\documentclass[ aip,jcp,amsmath,amssymb,reprint]{revtex4-1}

\usepackage{graphicx}
\usepackage{dcolumn}
\usepackage{bm}
\usepackage[T1]{fontenc}
\usepackage{lmodern}
\usepackage{epstopdf}
\usepackage{color}

\begin{document}

\title[Simulations of living filaments]
{Hybrid Molecular Dynamics simulations of living filaments}

\author{Mathieu Caby}
\author{Priscilla Hardas}
\author{Sanoop Ramachandran}
\author{Jean-Paul Ryckaert}
\email{jryckaer@ulb.ac.be}
\affiliation{ 
 Physique des Polym\`eres, Universit\'e Libre de Bruxelles,\\
Campus Plaine, CP 223, B-1050 Brussels, Belgium
}%

\date{\today}

\begin{abstract}
We propose a hybrid Molecular Dynamics/Multi-particle Collision Dynamics 
model to simulate a set of self-assembled semiflexible filaments and free 
monomers.
Further, we introduce a Monte-Carlo scheme to deal with single 
monomer addition (polymerization) or removal (depolymerization),
satisfying the detailed balance condition within a proper statistical
mechanical framework.
This model of filaments, based on the wormlike chain,  aims to represent 
equilibrium polymers with distinct reaction rates 
at both ends, such as self-assembled ADP-actin filaments in the absence 
of ATP hydrolysis and other proteins. 
We report the distribution of filament lengths and the 
corresponding dynamical fluctuations on an equilibrium trajectory.
Potential generalizations of this method to include irreversible steps like 
ATP-actin hydrolysis are discussed.
%
\end{abstract}

\maketitle
%
%
%
%
%
%
%
%
\section{Introduction\label{sec:intro}}

``Living'' polymers form a class of supramolecular structures characterized
by the reversible self-assembly of monomeric units into chains at 
equilibrium.~\cite{greer-02,schoot-05,cates-06,cates-90}
Some examples are systems forming 
discotic structures,~\cite{bushey-01,keizer-04}
fibrillar structures comprising of oligo($p$-phenylenevinylene) 
derivatives,~\cite{jonkheijm-06} 
and wormlike micelles.~\cite{rehage-91}
Filamentous structures of proteins found in the cell,
such as F-actin,~\cite{korn-87,pollard-09}
microtubules,~\cite{desai-97,flyvbjerg-94}
intermediate filaments~\cite{alberts} 
and bacterial pili~\cite{sauer-00,kline-10} are usually coined as 
non-equilibrium polymers.~\cite{howard} 
Their assembly/disassembly properties are coupled to irreversible chemical 
steps, like the ATP/GTP (adenosine triphosphate/guanosine triphosphate)
 hydrolysis to ADP/GDP  (adenosine diphosphate/guanosine diphosphate)
within the filament complexes which act as catalysts. 
These complexes containing hydrolyzed nucleotides turn out to be more 
loosely bounded within the filaments. 
This leads to a shortening of persistence length for the hydrolyzed 
portions and to different critical concentrations at active ends. 
In contrast to the passive polymerization of equilibrium polymers, 
the active polymerization/depolymerization confers to these polar 
biofilaments unusual properties such as treadmilling where the filament 
grows at one end and shrinks at the other,
or such as generation of work at the expense of chemical energy when a 
temporary network of filaments pushes on a membrane 
(e.g. in filopodia).~\cite{faix-06} 
Numerous in vitro experiments have been conducted to observe and analyze 
various peculiar properties of biofilaments. 
Tubulin fibers have been observed to alternate between slow growth and 
catastrophe shrinkage ~\cite{sneppen-zocchi} allowing for rapid 
reorganization of the cytoskeleton network while the role played by 
geometric boundaries on the growth dynamics of actin bundles was 
highlighted recently by Reymann et al.~\cite{reymann-10} 

There have been several theoretical investigations based on mean-field 
stochastic models to study the contour length and composition dynamics 
of single free actin (also microtubule) filaments in presence of ATP 
hydrolysis.~\cite{vavylonis-05,stukalin-06,ranjith-09} 
These theories have been adapted to investigate the behavior of bundles 
of filaments pushing against a flexible membrane~\cite{gholami-08} 
or wall~\cite{tsekouras-11} using straight and rigid filaments. 
They treat the free monomer concentration as homogeneous in space and 
constant in time and simplified hypotheses are assumed on the load 
distribution exerted by the surface on the filaments. 
Under conditions of confinement or crowding of bundles, it is expected 
that spatial correlations between filaments, their flexibility, the free 
monomer concentration fluctuations and the presence of local flows interfere 
with the main polymerization/depolymerization mechanism. 
To deal with more sophisticated situations where these aspects could be 
simultaneously taken into account, mesoscopic simulation approaches 
are needed. 
Pioneering Brownian Dynamics (BD) simulations, treating at the same 
time monomer diffusion and filament/network self-assembly, have been 
reported recently. 
They simulate actin networks pushing against an object~\cite{lee-08,lee-09} 
or a long single actin filament undergoing (de)polymerization steps in 
presence of ATP hydrolysis.~\cite{guo-09,guo-10}

In this paper, we propose and develop a mesoscopic model to simulate a set 
of self-assembled semiflexible filaments within a statistical mechanical 
framework. 
The total number of filaments and the total number of monomers are both 
fixed in the simulations, leading to semi-grand canonical ensemble at 
equilibrium. 
The filaments are modeled according to a discretized version of the 
wormlike-chain with a variable contour length and an adjustable persistence 
length. 
The chemical reactive steps are simulated via a novel Monte-Carlo (MC) move 
which forms the core of the present work. 
This MC move which satisfies the micro-reversibility conditions, is specifically 
suitable to model single addition/removal of a monomer at the 
active ends of the semiflexible filament. 
It is reminiscent of the MC moves developed for modeling scission/recombination 
kinetics in flexible linear micelles~\cite{huang-06,huang-09} or to model 
networks of reversible associating polymers~\cite{hoy-09} in which operational 
parameters can be fixed to separately adjust the equilibrium constant and the 
barrier height for any reversible chemical reaction. 
In absence of any additional irreversible step, we get a model for a set of 
equilibrium semiflexible filaments in a well defined ensemble. 
All these modeling aspects are detailed in Section~\ref{sec:gen-model}. 
Section~\ref{sec:theo} covers the theoretical treatment of this system, 
starting with an ideal mixture treatment and looking then for packing effects 
on the filament size distribution and on the kinetic rates. 
Our simulation experiments are detailed and their results analyzed in 
Sections ~\ref{sec:simpara} and~\ref{sec:results} respectively. 
In the Section~\ref{sec:discussion} , we come back on some methodological 
issues and envisage the extension to non-equilibrium polymers which requires the 
consideration of monomer flags to distinguish between hydrolyzed and 
non-hydrolyzed ATP complexes and the consideration of additional chemical steps. 
The application of our methodology to a specific biofilament is discussed by 
providing the explicit parametrization which would be needed to model 
ADP-F-actin (homopolymer of ADP-actin complexes in interaction with free 
ADP-actin monomers).
This  is an interesting example, not found in vivo, of a polar 
equilibrium polymer where the polymerization and depolymerization steps take
place at two kinetically inequal ends characterized by 
the same critical concentration. 
The paper closes with a short conclusion in Section~\ref{sec:conclusion}.

\section{General mesoscopic model of a set of interacting living 
filaments and free monomers in a thermal bath\label{sec:gen-model}}

We first describe the detailed microscopic model of the 
solute/solvent bath system. 
Our simulation method consists of four parts: the adopted solute model for a 
mixture of free monomers and semiflexible filaments of various 
lengths (Section~\ref{sec:solute}),
the solvent bath in which the solutes are immersed and the nature of 
the solute-solvent coupling (Section~\ref{sec:mpcd}),
an original stochastic (de)polymerization procedure which, together with 
the solute model, defines the ``living'' filaments concept 
(Section~\ref{sec:polydepoly}) and the statistical mechanics ensemble in 
which our finite-size system is sampled by our hybrid 
simulations (Section~\ref{sec:ensemble}).

\subsection{The solute \label{sec:solute}}

\begin{figure}
\centering
\includegraphics[scale=0.6]{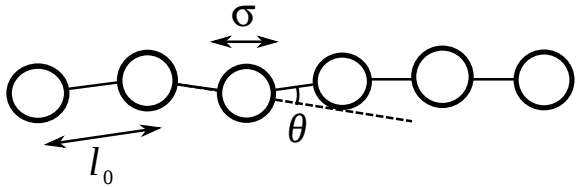}
\caption{ A schematic showing the model filament with beads of size $\sigma$ 
and bond length $l_0$.
The angle between successive bonds are denoted by $\theta$.
}
\label{fig:model}
\end{figure}

The solute system treated throughout this work is a set of $N_t$ spherical 
monomers of mass $m$. 
A fraction of these are present as free monomers. 
The remaining monomers constitute the building blocks of $N_f$ assembled 
polydisperse filaments with a range of sizes going from a minimum of three 
monomers to a maximum (set to an integer value $z$). 
The justification and the general consequences of imposed boundaries to the 
filament size will be discussed in Sections~\ref{sec:polydepoly} 
and~\ref{sec:ensemble}.

These filaments are modeled as semiflexible polymers with excluded volume 
interactions with its basic features adapted from a standard 
model.~\cite{ripoll-05,padding-09,hinczewski-09}
A schematic of the filament is shown in Fig.~\ref{fig:model}.
Any filament with $i$ monomers has $i-1$ bonding interactions 
$U_1(r)$ with an equilibrium length $l_0$ and an energy depth of 
$-\epsilon_0$
\begin{equation}
U_1(r)= \frac{1}{2} k(r-l_0)^2 
-\epsilon_0 \equiv U_1^{\rm vib}(r) + U_1^{\rm elec},
\label{eqn:U1}
\end{equation}
in terms of the absolute distance $r$ between adjacent monomers. 
The first purely vibrational term $U_1^{\rm vib}(r)$, where $k$ is the 
stretching force constant, maintains a chain structure with contour length 
$L_{\rm c}\simeq (i-1)l_0$. 
The second ``electronic'' term $-\epsilon_0$ corresponds to a 
(negative) bonding energy with respect to a reference zero energy level 
corresponding to the large distance non-covalent pair energy between monomers, 
in order to favor the polymerization step (see Section~\ref{sec:polydepoly}).

These non-covalent interactions are introduced via a shifted and truncated 
Lennard-Jones (LJ) potential, denoted by $U_2(r)$.
\begin{equation}
U_2(r)= 4\epsilon \left[
\left(\frac{\sigma}{r}\right)^{12}
-\left(\frac{\sigma}{r}\right)^{6}+
\frac{1}{4}
\right] \Theta(2^{1/6}\sigma-r). 
\label{eqn:U2}
\end{equation}
Here $\epsilon$ and $\sigma$ are the energy and diameter parameters of the 
LJ potential. 
The Heaviside step function satisfies
\begin{equation}
\Theta(x)=
\begin{cases}
0&\quad\text{for $x<0$},\\
1&\quad\text{for $x\geq0$}.
\end{cases}
\end{equation}
Both intramolecular (beyond second neighbors) and intermolecular pairs are 
subject to such excluded volume interactions.
Finally, a set of $i-2$ three-body potentials $\phi^{\rm bend}(\theta)$ 
is used to favor the straight alignment of successive bonds ($\theta=0$)
within each filament 
\begin{equation}
\phi^{\rm bend}(\theta) = 
k' (1-\cos\theta),
\label{eqn:potb}
\end{equation}
where $k'$ is the bending stiffness energy parameter, directly related to 
the persistence length $l_{\rm P}$ according to
$k'=k_{\rm B}T l_{\rm P}/l_0$. 
 
Note that in applications of the model to filaments, one typically has 
$l_{\rm P}\gg l_0$ and therefore, the short range (in physical distance) of 
the repulsive non-covalent interactions $U_2(r)$ implies that the latter 
will usually not operate as ``local interactions'' but possibly as 
``long-range interactions'' if the filaments are sufficiently long to 
behave as coils($L_{\rm c} \gg l_{\rm P}$).

The particles trajectories are obtained by solving the Newton's equations
using the velocity-Verlet integrator, as in any standard Molecular
Dynamics (MD) simulation.

\subsection{The solvent bath \label{sec:mpcd}}

The system of $N_t$ monomers is immersed in a sea of solvent particles 
of mass $m_s$ treated according to the Multi-Particle Collisional 
Dynamics (MPCD) model. 
This technique, first introduced by Malevanets and 
Kapral,~\cite{malevanets-99,malevanets-00} uses simplified solvent dynamics 
that faithfully reproduces the correct long time fluid behavior.
Subsequent to its introduction, it has been extensively studied 
theoretically as well as being used to investigate various complex systems
such as colloids, polymers and vesicles immersed in a 
solvent.~\cite{gompper-09}
Our use of this solute-solvent model, except for the living filaments aspects, 
is close to the semiflexible chains in solution.~\cite{ripoll-05,padding-09}

In this solvent model, all solvent particles move as free particles between 
(local) collisions taking place every $\Delta t$ time steps. 
The simulation volume $V$ is divided into cubic cells of side $a_0$. 
The local collisions which take place independently in each cell imply 
velocity changes of all particles of the cell 
(solvent particles, free monomers and monomers pertaining to a filament) 
while preserving the total mass, linear momentum and energy of that cell. 
Explicitly, within each cell, the relative velocity of each particle 
(with respect to the cell subsystem center of mass velocity) is rotated by a 
fixed angle $\alpha$ around a direction selected at random on a sphere with
uniform probability.

Between the collision steps, the solute dynamics is followed by solving the 
equations of motion of the solute subsystem, treated independently of the 
solvent. 
This coupling enables the proper thermalization of the solute particles 
as well as the incorporation of hydrodynamic effects.

\begin{figure}
\includegraphics[scale=0.6]{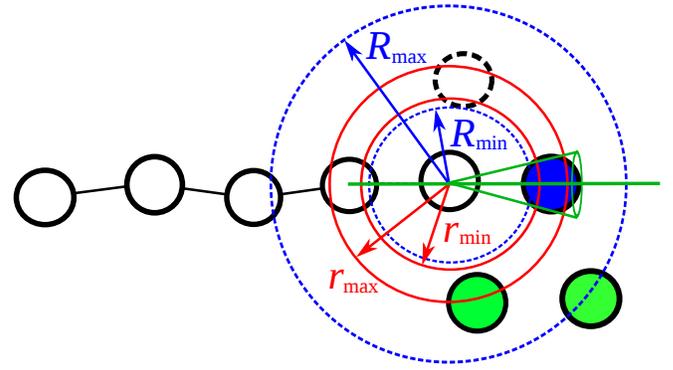}
\caption{\label{fig:stateI} 
State $I$: The polymerized state where the reacting monomer shown in dark blue 
is connected to the filament through intramolecular interactions 
and is, necessarily, located in a low intramolecular energy volume resulting 
from the intersection of a conic volume (in green) and the 
spherical layer (red) which respectively limit the bending and the stretching
 energies (see main text).
The dashed blue circles represent the volume $V_{\rm s}$
and the light green particles represent free monomers.
}
\end{figure}

\begin{figure}
\includegraphics[scale=0.6]{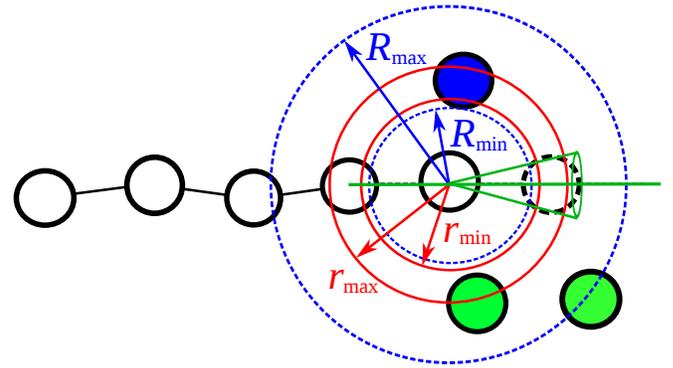}
\caption{\label{fig:stateJ} State $J$: The depolymerized state where the 
reacting monomer shown in blue is free but is necessarily located within the 
spherical layer shown with dashed-blue lines whose minimum radius limits 
the excluded volume pair interaction energy with the last monomer of the 
target filament end (see main text).
}
\end{figure}

\subsection{Modeling of (de)polymerization steps \label{sec:polydepoly}}

A fundamental property of bio-filaments such as actin, tubulin, pili etc., 
is their ability to grow/shrink via polymerization/depolymerization reactions 
at their ends.~\cite{alberts}
We propose to model these chemical reactions 
as instantaneous events which modify locally the topological 
connections of the so-called ``reacting monomer''. 
This change will be governed by a MC scheme, satisfying 
detailed balance, which is fully consistent with all the topological 
arrangements allowed by the semi-grand ensemble equilibrium partition 
function of monomers/filaments reacting mixture. 
Any instantaneous chemical reaction event taking place in our system will 
connect two microscopic states $I$ and $J$ differing only by the position of 
the single reacting monomer of (say index $i_1$) and by topological and 
geometrical constraints involving its link with the last monomer, 
(say of index $i_2$) of a particular active end of a filament and 
its immediate neighbor in the filament sequence (say index $i_3$). 
These three monomers which belong to the global set of $N_t$ solute monomers, 
are directly implied in the chemical step which we now describe with the help of 
Figs.~\ref{fig:stateI} and~\ref{fig:stateJ} respectively representing
the initial and final configurations of a polymerized or
depolymerized state.
In these figures, the dashed bead indicates the position of the unique 
reacting monomer $i_1$ in the alternative state.

In the polymerized state $I$ the reacting monomer, shaded dark blue
in Fig.~\ref{fig:stateI},  is linked to the last monomer $i_2$ of the
filament through a $U_1(r)$ bonding potential and contributes to the bending 
energy term $\phi^{\rm bend}(\theta)$ with monomers $i_2$ and $i_3$. 
It is further imposed that in state $I$, the polymerized monomer of index 
$i_1$ strictly lies within the region of low intramolecular energy volume 
$V_{\rm c}$ determined as the intersection of two volumes: 
i) the conical volume whose symmetry axis coincides with the bond vector 
joining monomers $i_3$ and $i_2$ with its tip located at monomer $i_2$ 
having a tip angle $\theta_{\rm max}$,
(ii) the spherical shell centered on the same monomer $i_2$ with inner and 
outer radii $r_{\rm min}$ and $r_{\rm max}$. 
This volume amounts to
\begin{equation}
V_{\rm c} = \frac{2\pi}{3}
(1-\cos\theta_{\rm max})
(r_{\rm max}^3-r_{\rm min}^3).
\label{eqn:Vc}
\end{equation}
The limiting values $r_{\rm min}$ and $r_{\rm max}$ are chosen such that 
$U_1^{\rm vib}(r_{\rm min}) = U_1^{\rm vib}(r_{\rm max}) = 3k_{\rm B}T$ 
and the extreme cone angle $\theta_{\rm max}$ is chosen such that 
$\phi^{\rm bend}(\theta_{\rm max})=3k_{\rm B}T$.
This implies that the volume $V_{\rm c}$ is the locus of points in space 
where the incremental increase in vibrational potential energy remains a few 
$k_{\rm B}T$.

In the depolymerized state $J$, where the reacting monomer $i_1$ is shown in 
dark blue, it is still located close to the filament end monomer 
$i_2$ but is now free (in the sense that it only interacts with the filament 
end through excluded volume interactions). 
More precisely, to be in the depolymerized reacting state $J$, the free 
reacting monomer $i_1$ must lie in a spherical layer centered on monomer 
$i_2$ with volume
\begin{equation}
V_{\rm s} = \frac{4\pi}{3}(R_{\rm max}^3-R_{\rm min}^3)
\label{eqn:Vs}
\end{equation}
where $R_{\rm min}$ is such that $U_2(R_{\rm min})=3k_{\rm B}T$ 
and $R_{\rm max}=2\:\times \:2^{1/6}\:\sigma $ 
(namely twice the range of the potential in Eq.~(\ref{eqn:U2})).

We now envisage the energy modifications involved by the transitions between 
$I$ and $J$ by focusing on the incremental energy of the reacting monomer 
$i_1$ with respect to the rest of the system. 
In state $I$, the total incremental energy, which contains both intramolecular 
and intermolecular contributions, can be written as
\begin{equation}
\epsilon^{{\rm tot}, I}_{i_1}=
\epsilon^{{\rm ev}, I}_{i_1}
+\epsilon^{{\rm vib}, I}_{i_1}-\epsilon_0,
\label{eq:epsi}
\end{equation}
where the contributions from the stretching energy $U_1^{\rm vib}(r)$ and the
bending energy $\phi^{\rm bend}$ involving the reacting monomer have been 
regrouped into $\epsilon^{\rm vib}$.
The sum of all excluded volume pair interactions $U_2(r)$ between monomer 
$i_1$ and the rest of the system is written as $\epsilon^{\rm ev}$.

The total incremental energy of monomer $i_1$ in state $J$ is the sum 
of all its excluded volume pair interactions with the rest of the system 
(at a different location in comparison with state $I$), giving
\begin{equation}
\epsilon^{{\rm tot}, J}_{i_1}=\epsilon^{{\rm ev}, J}_{i_1}.
\label{eq:epsj}
\end{equation}

Our chemical step model is best expressed in terms of reactive monomers. 
During the course of the simulation, any free monomer in the vicinity of the 
active end of a filament, i.e. located within the spherical layer of size 
$V_{\rm s}$ centered on the end-monomer of the filament, will be coined as 
``free reactive monomer'' meaning that it is susceptible to polymerize from 
a $J$-type depolymerized state. 
Note that the reactive end of a filament is active unless the filament has 
reached the maximum size of $z$ monomers, in which case it can only depolymerize.

Similarly, the last monomer at the active end of a filament containing a 
minimum of four monomers (trimers are not allowed to dissociate as filaments 
must have at least three monomers) will be considered as a reactive end-monomer,
 i.e. susceptible to depolymerize from an $I$-type polymerized state, 
as long as it is located within the low intramolecular energy volume 
$V_{\rm c}$.

During the time window in which a particular monomer is reactive, it can be the 
object of a chemical step.
The occurrence of such an event being sampled in a Poisson distribution of times 
with nominal frequencies fixed (for reasons which will be justified later) to
\begin{subequations}
\begin{align}
\nu^{\rm depol} \equiv \nu^{I\to J}&=
\nu \frac{V_{\rm s}}{V_{\rm s}+V_{\rm c}} \exp(-\beta \epsilon_0),
\label{eq:freqi}\\
\nu^{\rm pol} \equiv \nu^{J\to I}&=
\nu \frac{V_{\rm c}}{V_{\rm s}+V_{\rm c}}.
\label{eq:freqj}
\end{align}
\label{eq:freq}
\end{subequations}
The frequency prefactor $\nu$ is a free parameter which can be used to 
tune the trial frequencies (to tune the effective barrier height of the 
reaction) and hence the chemical rates in the system without affecting the 
equilibrium state.

When a chemical step involving a particular reactive monomer is selected 
by Poisson sampling statistics to occur at a specific time $t$, 
depending on the nature of the reactive monomer, an MC attempted move 
($I\to J$) or ($J\to I$) is made by sampling its new position uniformly 
from the volume $V_{\rm s}$ or $V_{\rm c}$ relative to the new $J$ or 
$I$ state. 
This new configuration is then accepted or rejected on the basis of an 
acceptance probability chosen to be
\begin{subequations}
\label{eq:accprob}
\begin{align}
P^{ I \to J}_{\rm acc} &=
{\rm Min}[1, \exp(-\beta(\epsilon^{{\rm ev}, J}_{i_1} 
-\epsilon^{{\rm ev}, I}_{i_1}-\epsilon^{{\rm vib}, I}_{i_1}))],
\label{eq:accprobi}\\
P^{ J \to I}_{\rm acc} &= 
{\rm Min}[1, \exp(-\beta(\epsilon^{{\rm ev}, I}_{i_1}
+\epsilon^{{\rm vib}, I}_{i_1}-\epsilon^{{\rm ev}, J}_{i_1}))],
\label{eq:accprobj}
\end{align}
\end{subequations}
where $\beta=(k_{\rm B}T)^{-1}$. 
In practice, an explicit sampling of this acceptance probability finally 
decides the success of the chemical step at the time $t$ originally decided 
from the Poisson samplings of reaction times. 
When the chemical step is accepted, the reacting monomer is instantaneously 
and definitively transferred to its new location and topological 
status while keeping its linear momentum unaltered. 
The dynamical trajectory is then continued for times greater than $t$ on the 
basis of the new microscopic state dynamical variables. 
On the contrary, if the attempted chemical step is not accepted, no reaction 
is recorded and the reactive monomer simply follows, with the rest of the 
system, the dynamical trajectory it would have followed in the absence of any 
chemical step at time $t$.

We note that finally, whether a chemical step is finally recorded or not, 
the reactive monomer remains reactive as long as it has not diffused out of the 
relevant $V_{\rm c}$ or $V_{\rm s}$  reactive volume in which it lies at 
times just later than $t$. 
This point illustrates the coupling between monomer diffusion and reactive 
events. 
It indicates that the effective kinetic rates will not be strictly proportional 
to $\nu$ at high frequencies as a too large value of the attempt frequencies 
will necessary lead to ineffective sequences of polymerization/depolymerization 
steps involving the same filament end and reactive monomer.

The justification of the above Poisson/MC algorithm to deal with 
(de)polymerization steps requires the proof that it satisfies detailed balance 
at equilibrium for any pair of specific microscopic states ($I$,$J$). 
This can be seen easily by considering the ratio of the number of transitions 
between the two states $N^{ I \to J}$ and $N^{ J \to I}$ which can be expressed 
as
\begin{equation}
\frac{N^{ I \to J}}{N^{ J \to I}}=
\frac{\exp{(-\beta \epsilon^{{\rm tot},I}_{i_1})} \nu^{ I\to J} 
(1/V_{\rm s}) P^{ I \to J}_{\rm acc}}
{\exp{(-\beta \epsilon^{{\rm tot},J}_{i_1})} \nu^{ J\to I} 
(1/V_{\rm c}) P^{ J \to I}_{\rm acc}}
\end{equation}
where both the numerator and the denominator contain four factors equal 
or proportional respectively to 
(i) the probability to be in the starting microscopic state $I$ or $J$, 
(ii) the probability per unit time to make an attempt starting from the 
original state towards the alternative state of the reactive monomer, 
(iii) the probability to reach the specific final microscopic state $J$ in 
$V_{\rm s}$ or $I$ in $V_{\rm c}$ in the attempted move and finally 
(iv) the probability to accept this final state, hence to record a successful 
chemical step. 
To verify that the above ratio is unity, one simply needs to envisage 
explicitly two cases, whether the combination 
$\epsilon^{{\rm ev}, I}_{i_1}+\epsilon^{{\rm vib}, I}_{i_1}
-\epsilon^{{\rm ev}, J}_{i_1}$ is positive or negative. 
Making it explicit for the former case, use of 
Eqs.~(\ref{eq:epsi}),~(\ref{eq:epsj}),~(\ref{eq:freq}) and
(\ref{eq:accprob}) leads to
\begin{equation}
\frac{N^{I \to J}}{N^{ J \to I}}=
\frac{\exp{(-\beta [\epsilon^{{\rm tot}, I}_{i_1}
-\epsilon^{{\rm tot}, J}_{i_1}])} \exp(-\beta \epsilon_0)}
 {\exp{(-\beta [\epsilon^{{\rm ev}, I}_{i_1}
+\epsilon^{{\rm vib}, I}_{i_1}
-\epsilon^{{\rm ev}, J}_{i_1}])}}=1
\end{equation}
where $\epsilon^{{\rm tot}, I}_{i_1}$ (or similarly for $J$) denotes the 
total incremental energy of the reacting monomer in state $I$. 
We note that alternative schemes are possible. 
However we observed that the present one maximizes the acceptance 
probabilities and thus optimizes the efficiency of the algorithm.

The MD algorithm is a step by step procedure where $h$ is the MD time step 
and $\Delta t$ (usually chosen such that $\Delta t=n_s h$ where $n_s$ is an 
integer) is the time interval between two successive stochastic collisions
in the MPCD procedure. 
The chemical reactions are integrated to the hybrid MD/MPCD scheme in the 
following way. 
At any discrete MD time $t$, we make an exhaustive list of reactive monomers.
This consists of filament-end monomers as well as free monomers which may
appear more than once (in the list) if they are active with respect to more 
than one filament end.
Then the occurence of each reactive step 
$(I \rightarrow J)$ or $(J \rightarrow I)$ during the next time interval between $t$ and $t+h$ is independently sampled by taking a random 
number $\zeta$ from a uniform distribution between $0$ and $1$ and compared 
to $1-\exp{(-\nu^{I\to J} h)}$ or to 
$1-\exp{(-\nu^{J\to I} h)}$.
Any detected occurence is recorded.
In most cases, after scanning the full list of potential reaction steps, we 
find zero or one occurence: in the latter case, the chemical step algorithm is
 applied at time $t$ until final MC acceptance or rejection of the new location 
and new topological link of the reactive monomer. 
In practice, we have never observed more than two occurrences during a MD step 
$h$ and in the rare double occurence cases, we attempt the 
reactions in succession.

\subsection{Simulation ensemble including chemical reactions 
\label{sec:ensemble}}

Theoretical developments and simulation experiments will be performed in 
the $(N_t, N_f, V, T)$ ensemble in which, besides the volume ($V$) and the 
temperature ($T$), the total number of monomers $N_t$ and the total number of 
filaments $N_f$ are both independently fixed. 
This ensemble is subjected to the additional constraint that the filament 
size, expressed in number of monomers, is restricted between a minimum of 
three and a maximum of $z$.
To any macroscopic state specified by a set of fixed thermodynamic variables, 
there will correspond an equilibrium free monomer number density $\rho_1$ and 
a filament length distribution $\{\rho_i\}_{i=3,z}$ which depends on the set 
of equilibrium constants $K_i$ associated to the relevant (de)polymerization 
steps connecting filaments of length $i$ and $i+1$.
Dynamic fluctuations at equilibrium involve various diffusive and reactive 
aspects which are in principle also species dependent.
Considerable simplification occurs if the equilibrium constant $K$ can be 
treated as filament size independent. 
The length distributions then become simple exponentials and the concept of 
(an overall) average polymerization rate, $U$, per active filament-end 
and the corresponding average depolymerization rate $W$ per active 
filament-end can be employed. 
In a subcritical regime ($U<W$) or in the supercritical regime ($U>W$), 
the length distribution must be respectively a decreasing or an increasing 
function of the filament's size, as a result of detailed balance. 
The ratio $U/W$ is equivalent to the ratio between the actual free monomer 
density $\rho_1$ and the state point dependent reference (``critical'') 
density, $\rho_{\rm c}=1/K$. 
This ratio $\hat{\rho}_1 \equiv \rho_1 / \rho_{\rm c}=U/W$ directly 
indicates whether the state point is subcritical ($\hat{\rho}_1 <1$) 
or supercritical ($\hat{\rho}_1 >1$). 
Within an ideal solution treatment, an $i$ independent equilibrium constant 
turns out to be a function of only the temperature, say $K^0(T)$, 
and a concentration independent critical density $\rho^0_{\rm c}=1/K^0$ 
more directly separates subcritical ($\rho_1<\rho_{\rm c}^0$) and 
a supercritical ($\rho_1>\rho_{\rm c}^0$) concentration regions. 

We close this section with some considerations about our choice to impose 
maximal and minimal bounds to the filaments in the context of living 
biofilaments, subject to single monomer polymerization or depolymerization. 
The presence of a lower bound to the size of existing filaments and the absence 
of consideration of any explicit nucleation mechanism imply that the number 
of filaments $N_f$ is strictly constant and fixed by initial conditions. 
As biofilaments are often generated by initiator proteins 
(like profilin for F-actin~\cite{pollard-03}) in vivo and in vitro biomimetic experiments, the 
imposition of a lower bound size in our simulations can be interpreted as a 
way to link the number of filaments to a fixed number of permanently active 
initiators. 
The imposition of an upper bound to the filament length follows primarily 
from our wish to treat living biofilaments in a simulation box for a 
broad range of equilibrium conditions, both in subcritical and supercritical 
conditions. 
In the former case, the upper-limit will have marginal effects if $z$ is 
large with respect to the characteristic size associated to the decaying 
exponential distribution. 
An equilibrated situation in supercritical conditions cannot be maintained 
with free living filaments as they would grow indefinitely. 
A stationary distribution is only possible if some confinement effect 
stops the unlimited growth of filaments. 
The $z$ upper limit should thus be seen as a simplified way to impose a 
confinement constraint allowing the establishment of the equilibrium state 
in supercritical conditions, a situation after all symmetric of the role of 
the lower bound (three monomers) in subcritical conditions. 
We finally note that in simulations, it will be pragmatic to adjust the size 
of the box $L$ such that the maximum length of the filaments satisfies 
$z l_0 < L$.

\section{Theoretical framework \label{sec:theo}}

We provide below some theoretical predictions on the filament length 
distribution and on the dynamics of filament length fluctuations for 
the equilibrium system at fixed temperature described in the previous section. 
Our system consists of living filaments with size restricted between lower 
and higher bounds, subject to single monomer (de)polymerization steps at 
imposed total monomer number density and fixed global filament number density. 
The theoretical framework provides the tools for a consistency check of our 
hybrid MD simulation technique by comparing in the next section the results 
of simulation experiments at various state points with the theoretical 
predictions. 
We start by approaching the filament distribution at a macroscopic level but 
rely to statistical mechanics to discuss the microscopic expressions of the 
equilibrium constants. 
The second part of this section deals with the time evolution of the filament 
length populations on the basis of kinetic equations adapted to our system and 
we provide the microscopic expressions of these chemical rates.

\subsection{The filament length distribution and equilibrium constants}

Restricting ourselves to equilibrium, we denote the free monomer density by 
$\rho_1$ and the filament densities by $(\rho_i)_{i=3,z}$. 
The constraints on the system are
\begin{equation}
\rho_1+ 3 \rho_3+ 4 \rho_4+...+z \rho_z-\rho_t=0,
\label{eq:rhotot}
\end{equation}
and
\begin{equation}
\rho_3+ \rho_4 +...+\rho_z-\rho_f=0,
\label{eq:rhoftot}
\end{equation}
where $\rho_t$ and $\rho_f$ are respectively the fixed monomer and filament 
global densities.

Chemical equilibrium requires that for each (de)polymerization reaction  
$A_i + A_1\rightleftharpoons A_{i+1}$ (implying filaments of length $i$ 
and $i+1$ and a monomer), the chemical potentials of the different species 
satisfy $\mu_{i+1}-\mu_{i}-\mu_{1}=0$ which can be transformed into an
expression relating densities 
\begin{equation}
\frac{\rho_i}{\rho_1 \rho_{i-1}} = K_i
\label{eq:Ki}
\end{equation}
where the equilibrium constant $K_i$ is a state point dependent quantity. 
Under the reasonable assumption that these equilibrium constants beyond the 
trimer ($i=3$) can be treated as independent of the filament size, the 
individual equilibrium densities can be expressed explicitly in terms of a 
unique constant $K$ which then turns out to carry the whole state point 
dependence of the distribution.

Applying $\rho_i=\rho_1 \rho_{i-1} K$ recursively and combining with 
Eqs. (\ref{eq:rhotot}) and (\ref{eq:rhoftot}), one gets an implicit expression 
linking the monomer density  $\rho_1$ and $K$,
\begin{equation}
\rho_t=\rho_1+\rho_f 
\frac{\sum_{i=3}^z i (K \rho_1)^{i-3}}
{\sum_{i=3}^z (K \rho_1)^{i-3}},
\label{eq:b1}
\end{equation}
and an expression for filament densities in terms of $\rho_1$ and $K$,
\begin{equation}
\rho_i= \rho_f \frac{(K \rho_1)^{i}}{\sum_{j=3}^z (K \rho_1)^{j}}.
\label{eq:Ni1}
\end{equation}

Analytical expressions corresponding to Eqs.~(\ref{eq:b1}) and (\ref{eq:Ni1}) 
can be obtained. 
Using reduced densities $\hat{\rho}_t=\rho_t K$, 
$\hat{\rho}_f=\rho_f K$ and $\hat{\rho}_i=\rho_i K$ to simplify expressions, 
we exploit the properties of finite geometrical series
$S_n \equiv  \sum_{k=0}^n (\hat{\rho}_1)^k 
= (1-(\hat{\rho}_1)^{n+1})(1-\hat{\rho}_1)^{-1}$ 
(note that $S_n$ is finite and continuous around
$\hat{\rho}_1=1$ for $n$ non negative and finite) and 
$T_n \equiv \sum_{k=1}^n k (\hat{\rho}_1)^k 
= \hat{\rho}_1 \partial S_n/\partial \hat{\rho}_1$. 
Equations (\ref{eq:b1}) and  (\ref{eq:Ni1}) can then be rewritten as
\begin{align}
\hat{\rho}_t
&=\hat{\rho}_1+\hat{\rho}_f \frac{3 S_{z-3}+ T_{z-3}}{S_{z-3}},
\nonumber\\
&=\hat{\rho}_1+\hat{\rho}_f 
\frac{3-2\hat{\rho}_1-(\hat{\rho}_1)^{z-2}[1+ z(1-\hat{\rho}_1)]}
{(1-\hat{\rho}_1)(1-(\hat{\rho}_1)^{z-2})},
\label{eq:rho1im}
\end{align}
and 
\begin{equation}
\hat{\rho}_i
= \hat{\rho}_f 
\frac{(1-\hat{\rho}_1)(\hat{\rho}_1)^{i-3}}{1-(\hat{\rho}_1)^{z-2}}
\equiv C \exp{(y\;(i-3))},
\label{eq:Ni2}
\end{equation}
where $y=\ln{\hat{\rho}_1}$ and 
$C=\rho_f (1-\hat{\rho}_1)[1-(\hat{\rho}_1)^{z-2}]^{-1}$. 

\begin{figure}
\centering
\includegraphics[scale=0.4]{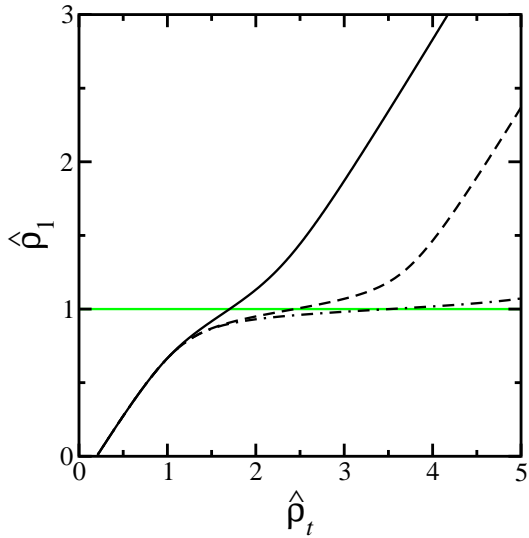}
\caption{$\hat{\rho}_1$ is plotted vs $\hat{\rho}_t$ with 
$\hat{\rho}_f=0.6699$ (fixed)
after numerical evaluation and inversion of 
Eq.~(\ref{eq:rho1im}). 
The solid line is for $z=18$,
the dashed line for $z=40$ and the dash-dotted line for $z=72$.
It is seen that when $z\to\infty$, $\hat{\rho}_1 \to 1$ as $\hat{\rho}_t$ 
becomes large.
The green (light) line at $\hat\rho_1=1$ is a guide for the eye. 
Such curves have only a universal character for ideal solution conditions 
as the reduction factor $1/K^0$ is only a function of temperature (see text).
}
\label{fig:b1bt-bound}
\end{figure}

The average size of the filaments in terms of monomers is then given by
\begin{equation}
N_{\rm av}
=\frac{\left[3-2 \hat{\rho}_1-\hat{\rho}_1^{z-2}(1+z [1-\hat{\rho}_1])\right]}
{(1-\hat{\rho}_1)(1-\hat{\rho}_1^{z-2})}.
\label{eq:nav}
\end{equation}
Let us stress that Eqs.~(\ref{eq:rho1im})-(\ref{eq:Ni2}), shown respectively 
in Figs. \ref{fig:b1bt-bound} and \ref{fig:pofi}, can alternatively be 
established via a statistical mechanics route (see Appendix~\ref{app:stat}). 
A semi-grand canonical ensemble with similar constraints takes into account 
all possible arrangements of $N_t$ monomers distributed into $N_f$ filaments 
and $N_1$ remaining free monomers, at fixed volume and temperature. 
Such an approach is richer as it allows the calculation of additional 
properties like the composition fluctuations in finite systems. 
Let us further point out that Eqs.~(\ref{eq:rho1im})-(\ref{eq:Ni2}) are not 
closed equations, unless $K$ can be established separately. 
The reduction factor ($1/K$) being generally state point dependent, the above 
relations are not universal curves. 
They merely represent consistency relationships between $\rho_1$, $\rho_i$ and 
$K$, for an arbitrary state point specified by ($T$, $\rho_f$, $\rho_t$).

The free monomer density $\rho_{\rm c}=1/K$ which corresponds to 
$\hat{\rho}_1=1$ is known as the critical density. 
If an explored state point in our ensemble accidently yields 
$\rho_1=\rho_{\rm c}$, all filament length densities 
are equal, as seen in Eq.~(\ref{eq:Ni2}), and $N_{\rm av}=(3+z)/2$. 
As discussed in the next subsection on the dynamics of the 
filament length, the polymerization and depolymerization steps are then 
equiprobable and each filament performs a one-dimensional (1D) discrete 
random walk bounded from below at $i=3$ and from above at $i=z$. 

For an arbitrary state point, $\hat{\rho}_1$ will differ from unity. 
If $\hat{\rho}_1>1$ (supercritical conditions), the filaments are subject to 
a similar but biased bounded random walk where polymerization steps are more 
frequent than depolymerization steps (except for $i=z$) and the stationary 
filament length distribution is an increasing function of the length in order 
to satisfy detailed balance. 
The opposite situation is observed for subcritical conditions, 
as $\hat{\rho}_1 < 1$. 
Figures~\ref{fig:b1bt-bound} 
and~\ref{fig:pofi}
illustrate respectively the equilibrium relationship Eq.~(\ref{eq:rho1im}) 
and the distribution Eq.~(\ref{eq:Ni2}). 
Equation (\ref{eq:rho1im}) for finite upper bound $z$ cannot easily be 
inverted and must be solved numerically.

\begin{figure}
\centering
\includegraphics[scale=0.4]{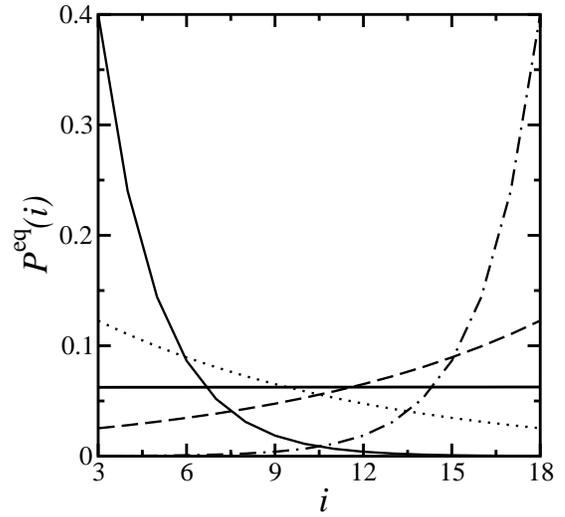}
\caption{
Theoretical bounded filaments distributions for subcritical 
($\hat{\rho}_1<1$) and supercritical ($\hat{\rho}_1>1$) monomer densities 
shown as $P^{\rm eq}(i)=\rho_i/\rho_f$ vs $i$. 
The solid, dotted, dashed and dash-dotted lines are 
for $\hat{\rho}_1=0.6,0.9,1.11$ and $1.67$ respectively in the 
conditions of the plot for $z=18$ in Fig.~\ref{fig:b1bt-bound}.
The solid horizontal line corresponds to the $\hat\rho_1=1$ case.
}
\label{fig:pofi}
\end{figure}

Provided $\hat{\rho}_1<1$, the $z \to \infty$ limit of the upper value of the 
filament range can be taken in Eqs.~(\ref{eq:rho1im}),~(\ref{eq:Ni2}) 
and~(\ref{eq:nav}) which simplify respectively to
\begin{align}
\hat{\rho}_t&=
\hat{\rho}_1+\hat{\rho}_f \frac{(3-2 \hat{\rho}_1)}{(1-\hat{\rho}_1)},
\label{eq:simpa}\\
\hat{\rho}_i &=\hat{\rho}_f (1-\hat{\rho}_1) (\hat{\rho}_1)^{i-3} 
\quad\mbox{with } i=3,\infty ,
\label{eq:Ninv} \\
N_{\rm av}&=3+\frac{\hat{\rho}_1}{(1-\hat{\rho}_1)}.
\label{eq:simp}
\end{align}
Here Eq.~(\ref{eq:simpa}) can now be inverted to yield
\begin{align}
\hat{\rho}_1&=\frac{1}{2}\Big[(1+\hat{\rho}_t-2 \hat{\rho}_f) 
\nonumber\\ 
&  -\sqrt{(1+\hat{\rho}_t-2 \hat{\rho}_f)^2-4(\hat{\rho}_t
-3 \hat{\rho}_f)}\;\Big].
\label{eq:inv}
\end{align}

All of the above expressions starting with Eq.~(\ref{eq:b1}), in particular the 
exponential distribution of filament sizes in Eq.~(\ref{eq:Ni2}), apply to 
equilibrium situations where the various chemical (de)polymerization steps are 
characterized by a unique (size independent) equilibrium constant $K$. 
Such a simplified macroscopic approach will be found to be fully relevant to 
our model system. 
Therefore, we now exploit standard statistical mechanics expressions of the 
equilibrium constants, first within an ideal gas approach and then within a 
dilute solution approach where interaction effects, between filaments and 
monomers, are included at the second virial coefficient level.

\subsubsection{Ideal solution}

We first consider an ideal solution case, reminiscent of the reacting ideal gas 
treated in standard books,~\cite{hill-thermody,mcquarrie}
where all interactions between filaments and free monomers are 
neglected, but where (de)polymerization steps nevertheless take place between 
the different (phantom) species, the filament + monomer solute system being 
coupled to a heat bath ``solvent'' at fixed temperature $T$. 
In that case, the equilibrium number densities $(\rho_1,\{\rho_i\}_{i=3,z})$ 
must satisfy the chemical equilibrium relationship
\begin{equation}
\frac{\rho_i}{\rho_1 \rho_{i-1}} = K_i^0
\label{eq:Ki0}
\end{equation}
where $K_i^0$, function of the temperature only, is related to single filament or 
single monomer canonical partition 
functions $q^0$ by~\cite{hill-thermody}
\begin{equation}
K_i^0=V \frac{q_i^0}{q_1^0 q_{i-1}^0}.
\label{eq:k0}
\end{equation}

The right hand side (RHS) of Eq.~(\ref{eq:k0}) can now be derived analytically 
for our filament model Hamiltonian provided only vibrational intramolecular 
interactions $U_1^{\rm vib}(r)$ and $\phi^{\rm bend}(\theta)$ are considered. 
As discussed in Section~\ref{sec:solute}, such a situation is valid as long as 
$L_{\rm c} \simeq l_{\rm P}> l_0$, because the intramolecular ``long-range'' 
monomer-monomer interactions which in fact act only at short physical distances 
cannot contribute for slightly bending rods. 
Using  Eqs.~(\ref{eqn:U1}) and~(\ref{eqn:potb}) for $i>2$, the kinetic 
contributions cancel out and thanks to the independence of the individual 
vibrational contributions, one finds an $i$ independent equilibrium constant 
for the RHS of  Eq.~(\ref{eq:k0}) to be 
\begin{align}
K^0(T)&=2\pi \exp(\beta\epsilon_0) 
\int_0^{\pi}
{\rm d}\theta 
~\sin\theta \exp{[- \beta \phi^{\rm bend}(\theta)]}
\nonumber\\
&\ \ \times 
\int_0^{\infty} 
{\rm d}r 
~r^2 \exp{[-\beta U_1^{\rm vib}(r)]} 
\label{eq:eqc0}\\
&=2\pi \exp(\beta\epsilon_0)
\frac{\left[ 1 - \exp(-2\beta k') \right]}
{\beta k'} \left(\frac{l_0}{x_0}\right)^{3} \nonumber\\
&\quad\times
\left[ \sqrt{\frac{\pi}{2}} \left(1+{\rm erf}[x_0]
\right)
(1+x_0^2) +x_0 \exp\left(-x_0^2/2\right) \right].
\label{eq:eqc}
\end{align}
Here $x_0\equiv (\beta k)^{1/2}l_0$, 
$\beta=(k_{\rm B}T)^{-1}$ and 
\begin{equation}
{\rm erf}[s]=\sqrt{2/\pi}\int_0^s {\rm d}t~\exp(-t^2).
\end{equation}
The derivation of Eq.~(\ref{eq:eqc}) from Eq.~(\ref{eq:eqc0}) is exact 
and relatively straightforward, requiring the use the integral 
\begin{equation}
\int_a^b{\rm d}u~u^2 \exp{(-u^2/2)}=
\left[\sqrt{\frac{\pi}{2}} {\rm erf}(u)-u \;\exp{(-u^2/2)}\right]_a^b.
\end{equation}
In our simulations, the values of the intramolecular parameters are chosen 
such that individual bond lengths have negligible fluctuations ($x_0\gg 1$) 
and filaments semiflexible ($L_{\rm c}\leq l_{\rm P}$). 
Therefore, Eq.~(\ref{eq:eqc}) of the equilibrium constant simplifies to
\begin{align}
K^0(T) &=
(2\pi)^{3/2} \beta^{-3/2} k^{-1/2} (k')^{-1}l_0^2
\exp(\beta\epsilon_0)
\label{eq:eqcs}
\end{align}
which more directly shows the impact of the model parameters on the equilibrium 
constant.

\subsubsection{Effects of interactions}

At this point, we introduce the activity coefficients $f_1$ and 
$\{f_i\}_{i=3,z}$ 
which measure for monomers and filaments of different lengths, the deviation 
from ideal behavior in the composition term of the chemical potential. 
In terms of these, the equilibrium constant associated to the reversible 
depolymerization of a filament of length $i$ can be expressed
as~\cite{hill-thermody} 
\begin{equation}
\frac{\rho_{i}}{\rho_1 \rho_{i-1}}
= \frac {f_1 f_{i-1}} {f_{i}}
K^0(T)
\equiv K_{i}.
\label{eqn:K}
\end{equation}
In Eq.~(\ref{eqn:K}), $K_i$ is a function of the thermodynamic state (and no 
longer a function of only the temperature). 
In presence of intermolecular forces, the $i$ dependence of $K_i$ for filaments 
is expected to be very weak while the magnitude of the intermolecular 
effects on $K_i$, for our model filament model, should primarily be related 
to the change in covolume resulting from the addition or the removal of a 
monomer at the end of a filament. 
This point can be analyzed quantitatively within the framework of the low 
density expansion of the activity coefficients. 
In Appendix~\ref{app:kik0}, we obtain a first-order estimate of $K_i/K^0$ as
\begin{align}
\frac{f_1 f_{i-1}}{f_i}&=
1 - \left[ 2 b_{11} + b_{1,i-1}- b_{1i}\right] \rho_1 \;- 
\nonumber\\
&\sum_{k=3}^z 
\left[b_{1k}+ (1+\delta_{k,i-1}) b_{k,i-1}- (1+\delta_{k,i}) b_{ki}\right]
 \rho_k \nonumber\\
&+ O(\rho^2)
\label{eq:expa}
\end{align}
where the $b$ factors are the expansion coefficients of the pressure 
in powers of the activities of the various species in the mixture. 
These factors can be expressed as follows
\begin{subequations}
\begin{align}
b_{11}&=\frac{1}{2V}(Z_{11}-V^2),\\
b_{ii}&=\frac{1}{2V}(Z_{ii}-V^2),\\
b_{1i}&=\frac{1}{V}(Z_{1i}-V^2),\\
b_{ij}&=\frac{1}{V}(Z_{ij}-V^2),
\end{align}
\label{eq:bz}
\end{subequations}
where the pairs of indices imply corresponding two body interactions and 
the pair configurational integrals $Z_{uv}$ between species $u$ and $v$ are 
defined as
\begin{align}
Z_{uv}&=\frac{Q_{uv} V^2}{q_u^0 q_v^0} \quad\quad \mbox{when } u \neq v,\\
Z_{uu}&=\frac{2 Q_{uu} V^2}{(q_u^0)^2}.
\end{align}
In these expressions, $q_u^0$ and $q_u^0$ are one body canonical partition 
functions and $Q_{uv}$ and $Q_{uu}$ are two body canonical partition functions 
for different and similar bodies respectively.

To be specific in this calculation of the second virial coefficients, we treat 
our filaments as rigid rods and we concentrate on the expressions relevant to 
the first order term in $\rho_1$ in the $K_i/K^0$ expression 
given by Eq.~(\ref{eq:expa}). 
We have~\cite{gray-gubbins}
\begin{align}
Z_{11}&= V \int {\rm d}\bm{r} \exp{(-\beta u_{11}(r))},\\
Z_{1i}&= V \int {\rm d}\bm{r} \exp{(-\beta u_{1i}(\bm{r},0,\bm{\omega}))},
\end{align}
where $u_{11}$ is the monomer-monomer pair potential and $u_{1i}$ is the 
intermolecular potential between a filament of length $i$ with center of 
mass at the origin and with (arbitrary) fixed orientation $\bm{\omega}$ and 
a free monomer located at $\bm{r}$. 
The explicit expressions of the corresponding $b$ factors are 
\begin{subequations}
\begin{align}
b_{11}&=
\frac{1}{2}
\int {\rm d}\bm{r} \left[\exp{(-\beta u_{11}(r))}-1 \right],\\
b_{1i}&
= \int {\rm d}\bm{r} \left[\exp{(-\beta u_{1i}(\bm{r},0,\bm{\omega}))}-1 \right].
\end{align}
\label{eq:bzint}
\end{subequations}
These integrals can be estimated by approximating the free monomers as hard 
spheres of diameter $\sigma$ and filaments as hard spherocylinders built as a 
cylinder of height $H=(i-1) l_0$ and radius $\sigma/2$, terminated by two 
hemispheres of radius $\sigma/2$. 
Performing the integrals gives
\begin{align}
b_{11}&=-V_{1}^{\rm cov}/2=-\frac{2}{3} \pi \sigma^3,\\
b_{1i}&=-V_{i}^{\rm cov}=-[\frac{4}{3}+(i-1)] \pi \sigma^2 l_0,
\end{align}
where $V_{1}^{\rm cov}$ and $V_{i}^{\rm cov}$ are the excluded volume of 
respectively a sphere or a spherocylinder, when approached by another sphere. 
Supposing that in Eq.~(\ref{eq:expa}) the monomer density is dominant with 
respect to the other terms involving filament densities, we get the net 
covolume effect on the equilibrium constant as,
\begin{equation}
K/K_0 \simeq 1+\frac{\pi}{3}\sigma^2 l_0 \rho_1.
\label{eq:Kth}
\end{equation}
This result gives at least the order of magnitude of intermolecular effects 
on the equilibrium constant and moreover suggests, as expected, that the 
$i$ dependent effects should be small.

\subsection{Mean field rate equations and rate constants}

\subsubsection{Effective (de)polymerization rate constant}

The (de)polymerization processes which take place within the system imply 
the following reaction
\begin{equation}
A_1+A_i \mathop{\rightleftharpoons}^{k_{\rm on}}_{k_{\rm off}}
A_{i+1},
\label{eqn:reacn}
\end{equation}
where $A_1$ represents a free monomer, $A_i$ a filament consisting
of $i$ monomers and $A_{i+1}$ one with $i+1$ monomers. 
In the mean field kinetics model, the $i$ independent $k_{\rm on}$ 
and $k_{\rm off}$ are respectively the polymerization and the depolymerization 
rate constants for filaments undergoing reactions at their active ends. 
Chemical equilibrium implies that the mean field equilibrium constant $K$, 
a function of the thermodynamic state, satisfies 
$K=k_{\rm on}/k_{\rm off}$.

Let us first consider polymerization events in an equilibrium situation with 
densities $(\rho_1,\{\rho_i\}_{i=3,z})$. 
The combination ${k_{\rm on}} \rho_1 \rho_i$ expresses the number of 
polymerization steps per unit of time and per unit volume involving 
specifically filaments of length $i$. 
In our general statistical mechanics model the filaments can only fluctuate 
between $i=3$ and $i=z$. 
Accordingly, the mean of the total number of reactive pairs 
for polymerization per unit volume (at any time) is given by 
$V^{*} \rho_1 (\rho_3+\rho_4+...+\rho_{z-1})$ where the volume factor equals to
\begin{equation}
V^{*}=4 \pi \int_{R_{\rm min}}^{R_{\rm max}} {\rm d}r~r^2g_{\rm em}(r)
\end{equation}
in terms of the equilibrium radial pair distribution $g_{\rm em}(r)$ between 
any filament active end and surrounding free monomers. 
In the ideal solution case, it simplifies to $V^{*}=V_{\rm s}$. 
According to our microscopic kinetic model, the total number of successful 
polymerization steps per unit volume and per unit time can be written as a 
product of three contributions, namely, 
the total number of reactive free monomers per unit volume, 
the polymerization attempt frequency  Eq.~(\ref{eq:freqj}), and finally, 
the polymerization acceptance probability Eq.~(\ref{eq:accprobj}). 
We thus have
\begin{align}
n^{\rm pol}=&k_{\rm on} \rho_1 (\rho_3+ \rho_4+..+ \rho_{z-1})= \nonumber\\
&V^{*} \rho_1 (\rho_3+ \rho_4+..+ \rho_{z-1}) 
   \times \nu \frac{V_{\rm c}}{V_{\rm s}+V_{\rm c}} \nonumber\\
& \times \langle {\rm Min}
[1,\exp{(-\beta ((\epsilon_n^{{\rm ev},I}+\epsilon_n^{{\rm vib},I})
-\epsilon_n^{{\rm ev},J}))}]
\rangle^{\rm co}.
 \label{eqn:npol}
\end{align}
Here the average $\langle ... \rangle^{\rm co}$ runs over all polymerizing 
events $J \to I$ sampled uniformly in the reacting volume 
$V_{\rm c}$.  
The incremental excluded volume energy of the reacting monomer $n$ in 
the different states $I$ or $J$ are represented by 
$\epsilon_n^{{\rm ev},I}$ or $\epsilon_n^{{\rm ev},J}$.
Also $\epsilon_n^{{\rm vib},I}$ is the 
non-negative incremental vibrational energy of the reacting monomer $n$ in 
state $I$ obtained by 
grouping the stretching potential $U_1^{\rm vib}(r)$ and the 
bending potential $\phi^{\rm bend}(\theta)$ provided by Eqs.(\ref{eqn:U1}) 
and~(\ref{eqn:potb}).

Now considering the depolymerization events at equilibrium, the average number 
of reactive (breaking) pairs per unit volume is given by 
$f^{*} (\rho_4+\rho_5+\rho_6+...+\rho_z)$ where the factor $f^{*}$ is the 
fraction of end monomers at the active end of the filaments which lie in the 
reactive volume $V_{\rm c}$. 

The total number of successful depolarization per unit of time and per unit 
of volume is, using depolymerization frequency Eq.~(\ref{eq:freqi}), 
and depolymerization acceptance probability Eq.~(\ref{eq:accprobi}), 
given by
\begin{align}
n^{\rm depol}&=
k_{\rm off} (\rho_4+ \rho_5+ \rho_6+..\rho_z) 
= \nonumber\\ 
&(\rho_4+ \rho_5+ \rho_6+..\rho_z) f^{*} \times 
\ \nu \frac{V_{\rm S}}{V_{\rm S}+V_{\rm c}}\exp(-\beta \epsilon_0) \nonumber\\ 
& \times \ \langle{\rm Min}
[1,\exp{(-\beta (\epsilon_n^{{\rm ev},J}-(\epsilon_n^{{\rm ev},I}
+\epsilon_n^{{\rm vib},I})))}]\rangle^{\rm sp}.
\label{eqn:ndepol}
\end{align}
In this case, the average $\langle ... \rangle^{\rm sp}$ runs over 
depolymerizing events $I\to J$ sampled uniformly in the spherical layer of 
volume $V_{\rm s}$.

To summarize, we finally get the rate expressions
\begin{align}
k_{\rm on} &=
\nu V^* \frac{V_{\rm c}}{V_{\rm c}+V_{\rm s}}\nonumber\\
&\quad\times \langle{\rm Min}
[1,\exp{(-\beta ((\epsilon_n^{{\rm ev},I}+\epsilon_n^{{\rm vib},I})
-\epsilon_n^{{\rm ev},J}))}]\rangle^{\rm co},\\
k_{\rm off} &=
\nu f^* \frac{V_{\rm s}}{V_{\rm c}+V_{\rm s}}
\exp(-\beta\epsilon_0) \nonumber\\
&\quad\times \langle{\rm Min}
[1,\exp{(-\beta (\epsilon_n^{{\rm ev},J}-(\epsilon_n^{{\rm ev},I}
+\epsilon_n^{{\rm vib},I})))}]\rangle^{\rm sp}.
\end{align}
The order of magnitude of these rates for reasonably dilute solutions 
can be estimated by neglecting excluded volume interactions, 
using $f^{*} \simeq 1$, $V_{\rm c} \ll V_{\rm S}$ and the ideal solution 
equilibrium constant $K^0$ given in Eq.~(\ref{eq:eqcs}), as
\begin{subequations}
\label{eq:konkoffest}
\begin{align}
k_{\rm off}^{\rm estim} & \simeq \nu \exp(-\beta\epsilon_0),
\label{eqn:konkoffest-a}\\
k_{\rm on}^{\rm estim} & \simeq K^0 \nu \exp(-\beta\epsilon_0).
\end{align}
\end{subequations}

\subsubsection{Mean-field filament length dynamics}

The filament length dynamical evolution is usually treated theoretically in 
terms of a chemical master equation with the reaction rates treated as 
input constants. 
For biological filaments like actin, 
filaments change by one unit 
and, if we disregard for the present time the existence of variants in the 
complex formation of the monomers and the modification of ATP 
into ADP through hydrolysis, only the polymerization rate $U$ and 
the depolymerization rate $W$ are relevant for the single monomeric species 
considered. 
In our case where the filament lengths fluctuate between 
$i=3$ and $i=z$, the set of equations governing population dynamics reads
\begin{subequations}
\label{eq:master}
\begin{align}
\frac{{\rm d}P(3,t)}{{\rm d}t}&= -U P(3,t) + W P(4,t)\label{eq:master2}\\
\frac{{\rm d}P(k,t)}{{\rm d}t}&= -(U+W) P(k,t) + U P(k-1,t) \nonumber \\
&\ \ \ \ + W P(k+1,t)\label{eq:master1}\\
\frac{{\rm d}P(z,t)}{{\rm d}t}&= -W P(z,t) + U P(z-1,t)
\end{align}
\end{subequations}
where $P(i,t)$ is the probability to observe filaments with length $i$ 
at time $t$ and in Eq.~(\ref{eq:master1}) the index $k$ runs from 
$4$ up to $z-1$. 
This set of equations is solved for constant rates $U$ and $W$ for 
given initial conditions $P(i,0)$.
This approach which treats independent filaments is physically justified if the 
monomer density (apparently uncoupled to the dynamical scheme) is assumed to be 
homogeneous in space and in time. 
Well stirred systems with compensating external addition or removal of free 
monomers provide plausible non-equilibrium stationary conditions for which the 
set of Eqs.~(\ref{eq:master}) is pertinent.

In our context we explore different specific values of 
$U=k_{\rm on} \rho_1$ and $W=k_{\rm off}$. 
The Eqs.~(\ref{eq:master}) can be tested by our simulations if we probe 
the dynamical filament length fluctuations at equilibrium. 
For example, by picking up a subset of filament of length $j$ at some time 
$t=0$ and following the time evolution of their size distribution, one should 
get an evolution from the original delta function back to the equilibrium 
distribution. 
This distribution is simply the conditional probability $P(j,t;i)$ that a 
filament having a length $i$ at an initial time $t=0$ has a length $j$ 
at time $t$.

This conditional probability time evolution will reflect the statistical 
properties of the bounded biased random walk induced by the chemical steps. 
However, if the starting size $i$ is not immediately close to a boundary at 
$i=3$ or at $i=z$, the behavior of $P(j,t;i)$ will be representative of an 
unbounded biased random walk as long as the measuring time $t$ remains much 
smaller than the (shortest) mean first passage time towards one of the 
boundaries. 
At long times, boundaries have a major influence (in any subcritical or 
supercritical regime) on the re-establishment of the equilibrium distribution.

\begin{table*}
\caption{\label{tab:res}
Table with simulation results for different values of $N_t$.
The particle packing fraction is represented by $\eta$.
We fix $N_f = 80$ in all the experiments with their size limited
between 3 and $z=18$.
Here $\hat\rho_1^{\rm sim}$ is the reduced monomer density obtained
from the slope of the logarithm of the equilibrium filament length distribution,
and $\rho_1^{\rm sim}$ is the measured single monomer density.
For comparison, we also tabulate $\rho_1$ numerically calculated from 
Eq.~(\ref{eq:rho1im}).
The polymerization ($U$) and depolymerization ($W$) rates  
are calculated directly by counting.
}
\begin{ruledtabular}
\begin{tabular}{c c c c c c c c c c}
Exp.&
$N_{\rm t}$&
$\eta$&
$\hat\rho^{\rm sim}_1$&
$\rho_1^{\rm sim}$&
$\rho_1$ 
&
$U$ &
$W=k_{\rm off}$ &
$k_{\rm on}=U/\rho_1^{\rm sim}$ \\
\hline
& & & & & & & &\\
G1 &				
1076&  				
0.0121&  				
0.60658	$\pm$ 0.004&		
0.12203	$\pm$ 0.0002& 		
0.12227&  			
0.00244&  			
0.00399&  			
0.02000  			
\\
G2 &				
1744&  				
0.0196&  				
0.91490	$\pm$ 0.003&		
0.17994	$\pm$ 0.0006& 		
0.18009&  			
0.00356&  			
0.00389&  			
0.01978  			
\\
G3 &				
2008&  				
0.0225&  				
1.00756	$\pm$ 0.001&		
0.19805	$\pm$ 0.0003& 		
0.19808&  			
0.00388&  			
0.00385&  			
0.01959  			
\\
G4 &				
2338&  				
0.0262&  				
1.13775	$\pm$ 0.001&		
0.22141	$\pm$ 0.0003& 		
0.22168&  			
0.00432&  			
0.00379&  			
0.01951  			
\\
G5 &				
3311&  				
0.0372&  				
1.78321	$\pm$ 0.008&		
0.33824	$\pm$ 0.0002& 		
0.33831&  			
0.00637&  			
0.00360&  			
0.01883  			
\\
G6 &				
4000&  				
0.0449&  				
2.39206	$\pm$ 0.020&		
0.44891	$\pm$ 0.0001& 		
0.47836&  			
0.00821&  			
0.00345&  			
0.01829  			
\\
\end{tabular}
\end{ruledtabular}
\end{table*}

\section{Adopted microscopic parameters and list of simulation experiments. 
\label{sec:simpara}}

All the physical quantities will be expressed in a system of units based on: 
the length $a_0$ of the side of the cell in MPCD, the energy $k_{\rm B}T$ 
(where $k_{\rm B}$ is the Boltzmann constant) and the mass $m_s$ of a solvent 
particle.
The resulting time unit is expressed as $\tau_0= a_0 \sqrt{m_s/(k_{\rm B}T)}$.

For the monomer/filament model, we adopt a unique set of parameters inspired 
by previous works on semiflexible chains in MPCD 
solvent~\cite{ripoll-05,padding-09} or in BD 
solvent~\cite{hinczewski-09}, namely beads of mass $m=5.0$ with size 
$\sigma=0.44545$ 
and energy parameter $\epsilon=3$. 
Within the filament, monomers are linked by hard springs with equilibrium length 
$l_0=0.5$ and force constant $k=1600$. 
The bending parameter is set to $k'=20$, which implies a persistence length of 
$k'/(k_{\rm B}T)=l_{\rm P}/l_0=20$. 
These values lead to a reactive volume $V_{\rm s}=3.81855$ 
(see Eq.~(\ref{eqn:Vs})) defined by $R_{\rm min}= 0.44545$ 
(for which the potential is $3k_{\rm B}T$), and 
$R_{\rm max}=1$, while those defining the reactive volume 
$V_{\rm c}=2.9 \times 10^{-2}$ (see Eq.~(\ref{eqn:Vc})) are given by 
$r_{\rm min}= 0.43876$, $r_{\rm max}= 0.56124$ and 
$\theta_{\rm max}= 0.55481$ rad.

The MPCD parameters are chosen to match a well studied state 
point,~\cite{ripoll-05} 
given by a solvent density of $\rho_{\rm solv}=5$, 
box collision angle of $130^{\circ}$ and 
a collision frequency $1/\Delta t=10$. 
At this solvent state point, it is known that a particle of mass 
$m=5m_s$ interacting at infinite dilution via MPCD collisions with solvent, 
has a diffusion coefficient~\cite{ripoll-05} of $D_0=0.043$ so that an estimate 
of the diffusion time is $\tau_{\rm D}=(0.463428)^2/D_0=5.0$ 
(where $r=0.463428$ is the distance where the potential $U_2(r)= k_{\rm B}T$).

With the adopted kinetic parameters $\nu=5.0$ and $\epsilon_0=6.9$ one gets at 
$k_{\rm B}T=1$, the ideal solvent equilibrium constant $K^0=4.88373$.
This allows the estimates of rates, from Eqs.~(\ref{eq:konkoffest}), as 
$k_{\rm off}^{\rm estim} =5.\times 10^{-3}$ and 
$k_{\rm on}^{\rm estim}=2.5\times 10^{-2}$ respectively.

All the simulation experiments, conducted at a common temperature 
$k_{\rm B}T=1$, 
follow $N_f=80$ filaments with size fluctuating between three and $z=18$ 
which are enclosed in a cubic volume $V=18^3$ containing $29,160$ MPCD solvent 
particles. 
Note that the box length ($L=18$) is twice the maximum allowed filament length. 
Periodic boundary conditions are applied in all three dimensions. 
Table~\ref{tab:res} reports the total number $N_t$ of monomers which 
is specific to each experiment, together with experiment codes
G1 - G6 
increasing with $N_t$ and the associated volume packing fraction $\eta$. 
In the simulations, the MD integration time step $h$, associated to the Verlet 
algorithm, has been set to $h = 0.002$, which corresponds exactly to 
$\Delta t/50$.

A typical simulation setup involves an equilibration run of $5 \times 10^{5}$ 
corresponding to $\approx 2.5\times 10^3$ depolymerization steps per active 
filament end or to $\approx 10^5 \tau_D$. 
Four runs of time duration of $10^5$ are then performed in succession to get 
four independent estimates of the averages of interest which are then used 
to determine the global average (over a total time of 
$t_{\rm exp}=4.\times 10^5$) and an estimate of the associated statistical 
error.

\section{Simulation results \label{sec:results}}

\begin{figure}
\centering
\includegraphics[scale=0.4]{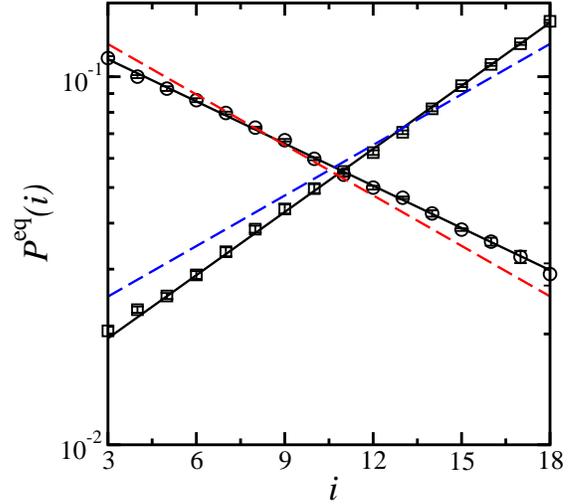}
\caption{
Plot of $P^{\rm eq}(i)=\rho_i/\rho_f$ vs $i$ for two different 
$\hat\rho_1$ values, namely subcritical experiment G2 (black circles) 
and supercritical experiment G4 (black squares). 
The solid black lines are the corresponding fits.
The two dashed straight lines indicate the expected distributions in ideal 
solution conditions for the same state point ($N_t, N_f, V, T$).}
\label{fig:dist}
\end{figure}

\subsection{Equilibrium distributions \label{sec:eqresults}}

The experiments differing in the total number of monomers $N_t$ in the box lead 
to a resulting free monomer density $\rho_1^{\rm sim}$ (see Table~\ref{tab:res}) 
and a distribution of filament densities for all accessible lengths (monomer 
number). 
In all cases, the distribution of filament lengths turns out to be exponential, 
confirming the hypothesis of a unique (state point dependent) equilibrium 
constant $K$ for all filament lengths. 
In Fig.~\ref{fig:dist}, $P^{\rm eq}(i)\equiv \rho_i/\rho_f$ (in log scale) 
versus $i$ (for two cases) is plotted.
It shows a linear behavior with a slope $\alpha$ providing directly the reduced 
free monomer density $\hat{\rho}_1 \equiv \rho_1 K = \exp{(\alpha)}$. 
All the $\hat{\rho}_1$ data obtained from the filament distributions of the 
various experiments are gathered in Table~\ref{tab:res}. 
It shows that as $N_t$ increases, the system goes from subcritical to 
supercritical conditions, with the experiment G3 being very close to the 
critical situation. 
In Fig.~\ref{fig:dist}, we also show the distribution expected (dashed lines)
for the same state point ($N_t, N_f, V, T$) if ideal solution conditions were 
applicable with equilibrium constant $K^0$ computed from Eq.~(\ref{eq:eqcs}).

\begin{figure}
\centering
\includegraphics[scale=0.4]{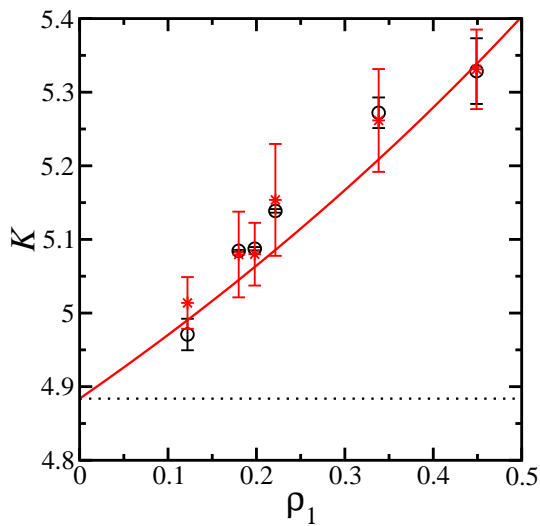}
\caption{ 
The equilibrium constant calculated from the simulations.
The open black circles are calculated from the free monomer density and the 
exponential filaments densities while the red stars are estimated from the 
ratio of the rates $(1/\rho_1) U/W$. 
Within the error bars, the agreement between the two estimates is excellent. 
The solid red line is the ratio $(1/\rho_1) f(\rho_1)/g(\rho_1)$ estimated from 
the fits.
The dotted line represents the equilibrium constant for the ideal 
case.
}
\label{fig:kvsa1}
\end{figure}

By dividing $\hat{\rho}_1$ by $\rho_1$ for each line of the Table~\ref{tab:res}, 
one gets a first estimate of the equilibrium constant $K$ which is shown in 
Fig.~\ref{fig:kvsa1}, as a function of $\rho_1$ itself. 
We postpone the discussion of the increase of $K$ versus $\rho_1$ until we 
confirm these data with a second estimate of $K$ extracted from the number of 
chemical (de)polymerization steps in the next subsection.

\subsection{State point dependence of rates and related equilibrium constant 
\label{sec:dyna}}

\begin{figure}
\centering
\includegraphics[scale=0.4]{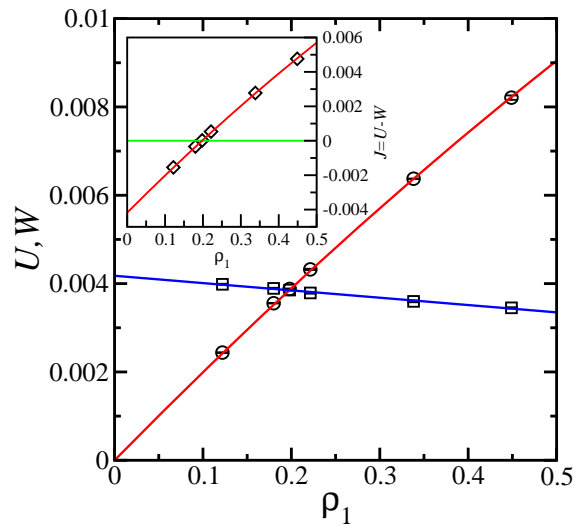}
\caption{Polymerization rate $U$ (circles) and depolymerization rate $W$ 
(squares) vs $\rho_1$ for all experiments. 
The solid lines are linear ($W$) and quadratic ($U$) fitting curves 
discussed in the text.
In the inset, the global growth rate $J=U-W$ is indicated (diamonds).
The solid line showing $J=0$ is only for reference.
}
\label{fig:uw}
\end{figure}

The effective rates $W=k_{\rm off}$ and $U=k_{\rm on}\rho_1$ can be estimated, 
on the basis of the assumption that they are independent 
of filament length, using expressions in Eqs.~(\ref{eqn:npol}) 
and~(\ref{eqn:ndepol}). 
The quantities $n^{\rm pol}$ and $n^{\rm depol}$ in these expressions are 
obtained by the counting of the total number of successful polymerization and 
depolymerization steps observed during the simulation, divided by the volume 
$V$ and the total simulation time.
The rates for the six experiments are gathered in the Table~\ref{tab:res} and 
are also reported in the Fig.~\ref{fig:uw}.

We observe that $W$ slightly decreases with increasing $\rho_1$,
a property which must be related to a decrease in the rate of acceptance of an 
attempted depolymerization step due to higher packing density.
Similarly, the polymerization rate is not strictly linear in $\rho_1$ and 
a negative quadratic term can be extracted from a fit. 
We perform the three parameters combined fit 
$W(\rho_1)\equiv f(x)=a x + b$ and 
$U(\rho_1)=k_{\rm on} \rho_1 \equiv g(x) = c x^2 + K^0 b x$ 
as we impose that for $\rho_1\to 0$, the ratio of $k_{\rm on}$ and $k_{\rm off}$ 
reproduce the thermodynamic value of $K^0= 4.88373$. 
We find $a = - 0.00165663$, $b =   0.00417753$ and 
$c = - 0.00461497$. 
To first order, it gives the relationship 
$K/K^0  =   1 + A x$ with $A = c(bK^0)^{-1} - ab^{-1}= 0.169$. 
This value can be compared with the theoretical prediction given by 
Eq.~(\ref{eq:Kth}), which gives $A'= 0.112$ if the size of the monomer is 
taken to be equal to the distance at which the energy is of the order of unity
($\sigma \approx 0.463$). 
The estimate has the correct order of magnitude but is about $35\%$ too low 
perhaps because the covolume effects of the filaments are included in the fitted 
quantities ($U$ and $W$) but are absent in the theory. 
We believe however that the main source of increase of the equilibrium constant 
with $\rho_1$ finds its origin in the covolume effect described in 
Eq.~(\ref{eq:Kth}).

\subsection{Dynamic fluctuations of filament length \label{sec:ldist}}

\begin{figure}
\centering
\includegraphics[scale=0.4]{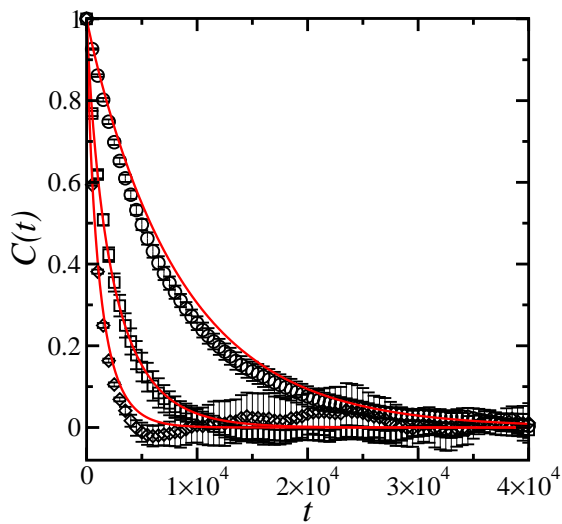}
\caption{ $C(t)$ vs $t$ for experiments G1 (squares), 
G3 (circles) and G5 (diamonds).
The solid curves represent the theoretical expressions of $C(t)$ based on 
Eqs.~(\ref{eqn:nn}) and~(\ref{eqn:ct}) where the conditional probabilities are 
computed by solving according to standard methods, the mean field 
population dynamics given by Eqs.~(\ref{eq:master}), 
using the $U$ and $W$ actual tabulated values of the same state point.
}
\label{fig:ctvst}
\end{figure}

At equilibrium, we consider the (static) population probability 
$P^{\rm eq}(i=3,z$) and the conditional probability $P(j,t;i)$ for a 
filament to have a length $j$ at time $t$ if its length was $i$ at time 
$t=0$. 
In terms of these, one can express the time correlation function of filament 
length $N$ by
\begin{equation}
\langle N(0) N(t)\rangle=\sum_{i=3}^z \sum_{j=3}^z i\;j\;P^{\rm eq}(i)P(j,t;i).
\label{eqn:nn}
\end{equation}
The normalized correlation function of the length fluctuations around their 
mean is formally
\begin{equation}
C(t)=\frac{\langle N(0)N(t)\rangle-\langle N\rangle^2}
{\langle N^2\rangle -\langle N\rangle^2}
\label{eqn:ct}
\end{equation}
where $\langle N\rangle=N_{\rm av}$ and 
$\langle N^2\rangle=\sum_{i=3}^z i^2 P^{\rm eq}(i)$. 
Figure~\ref{fig:ctvst} shows the variation in $C(t)$ with $t$ for 
the experiments G1, G3 and G5.
These curves are in excellent agreement with the theoretical expressions of 
$C(t)$ based on Eqs.~(\ref{eqn:nn}) and~(\ref{eqn:ct}) where the conditional 
probabilities are computed by solving the mean field population dynamics 
Eqs.~(\ref{eq:master}), using the tabulated values of $U$ and $W$ 
at the same state point. 
We observe that the relaxation time of these fluctuations is maximum at critical 
density (experiment G3), as a result of the uniform distribution of filament 
lengths which requires a filament to explore a wide range of lengths before 
loosing memory of its initial value.

\begin{figure}
\centering
\includegraphics[scale=0.4]{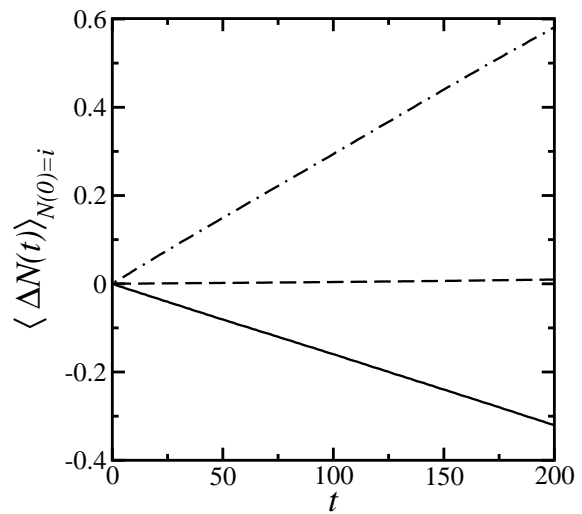}
\caption{ 
$\langle \Delta N(t)\rangle_{N(0)=i}$ vs $t$ at short times for different 
values of $\hat\rho_1$. 
The lines refer (from bottom) to an average over filaments $i=8,9$ and 
$10$ in experiment G1, an average over filaments 
$i=9,10$ and $11$ in experiment G3 and an average over filaments 
$i=11,12$ and $13$ in experiment G5.
}
\label{fig:dlvst}
\end{figure}

In Fig.~\ref{fig:dlvst}, we observe the growth or decay of the length of 
filaments in various super or subcritical regimes, focusing on the short time 
behavior of the conditional probabilities for particular filament lengths 
$i$ which are present in reasonable number and located 
relatively far from the boundaries at three or $z$ at the same time zero.
The quantity measured and shown in Fig.~\ref{fig:dlvst} is
\begin{equation}
\langle\Delta N(t)\rangle_{N(0)={i}}=\sum_{j=3}^z \:j\;P(j,t;i)-i,
\end{equation}
and is also averaged over a few appropriate filament lengths. 
Since the largest (de)polymerization rates are of the order of $0.01$, 
it can be expected that for times  $t_{\rm obs}= 100\sim200$, 
the filament populations of $P(j,t;i)$ for $j=3$ or $j=z$ are still negligible 
and the behavior of $\langle \Delta N(t)\rangle$ is representative of a 
situation at the same density $\rho_1$ of unbounded filaments. 
We have verified that the slopes of the straight lines observed in 
Fig.~\ref{fig:dlvst} are in full agreement with the global growth or decay 
rate $J$ shown in the inset of Fig.~\ref{fig:uw}.

\section{Discussion
\label{sec:discussion}}

The methodology presented in this work applies to the sampling of a reactive 
(semi-grand) canonical ensemble consisting of a mixture of free monomers and a 
fixed number of self-assembled linear polymeric filaments. 
Specifically, the filaments undergo reversible single monomer
end-filament polymerization/depolymerizations at one or both active ends of 
any filament.
Regarding the filament model, a discretized wormlike chain is used as the 
basis of an extension towards a ``living wormlike chain'' version where the 
contour length is variable. 
In this section, we mention some specific details of our method, some of them 
with respect to alternative approaches proposed in the literature. 
We briefly indicate how our methodology can be generalized to include additional 
or alternative kinetic processes found in biofilaments such as filament rupture, 
distinction between ADP and ATP complexes and irreversible hydrolysis processes. 
We organize this discussion by regrouping some purely 
methodological aspects first, and then consider the explicit case of actin to 
illustrate the versatility of our approach.

\subsection{Methodological aspects}

The dynamic trajectory of our solute (reactive mixture of filaments and free 
monomers), coupled to a bath of solvent particles, is obtained via a numerical 
scheme which incorporates MC reactive steps in the hybrid MD-MPCD 
method. 
For non-reacting fluids, the hybrid algorithm conserves the total energy but 
in our case, the MC chemical steps give an overall canonical ensemble 
character to the sampling. 
The imposed canonical temperature $T^{\rm can}$ is set in the acceptance 
probabilities of the chemical step Eqs.~(\ref{eq:accprob}) and in the attempt 
depolymerization frequency Eq.~(\ref{eq:freqi})). 
Between two successive chemical events, the system's total energy is conserved 
but it is discontinuous when a chemical step is successful. 
Pragmatically, the initial equilibration run to probe a new state point is 
started with initial velocities for solute monomers and solvent particles 
such that the average kinetic energy per degree of freedom is set equal to 
$k_{\rm B}T^{\rm can}/2$. 
As the system equilibrates while undergoing chemical steps at the same 
prescribed temperature, the instantaneous average kinetic energy per degree of 
freedom fluctuates in time around the expected stable value. 
The choice of coupling of the new chemical steps proposed 
in this paper to a MPCD solvent may be questioned.
This (continuous space) particulate description of the solvent locally conserves 
momentum and hence preserves long wavelength fluid hydrodynamics as well as
allows the  boundary conditions to be easily adjusted. 
This may play an important role in highly confined geometries such as in 
biofilaments in a microchannel or actin bundle developments in filopodia.

The stochastic scheme to model the (de)polymerization steps is a 
purely instantaneous local perturbation. 
This is meant to represent at a mesoscopic level (coarse graining in space and 
time), the direct coupling effect of many internal and external degrees of 
freedom of the reacting pair ``filament end''-``protein complex'' which are 
eliminated.
The modeling of the chemical step introduces a local discontinuity in the 
position of the reacting monomer but not in the monomer velocity. 
This is the price to be paid in order to succeed in devising a scheme 
in which the free reacting particle can be effectively moved to/from a position 
in space where it participates with the other degrees of the filament to an 
equilibrated intramolecular configuration. 
In real systems, as one protein binds to an existing protein filament in 
solution, it requires some time to modify its internal structure, 
to adjust its orientation and its translational location 
so that intermolecular interactions between units stabilize the newcomer. 
This waiting time corresponds in our model to the sampled reaction time in the 
Poisson process governed by the attempt frequency in
Eqs.~(\ref{eq:freq}), which can be tuned via the free parameter $\nu$.
The reactive steps are modeled as local events obeying micro-reversibility. 
They are effectively coupled to the solute degrees of freedom through structural 
properties like the filament end-monomer/free-monomer pair correlation function, 
and through the various filament and free monomer diffusive processes. 
Our simulation results indicate a weak variation of the rates with state point 
at moderate solute packing fractions when the effective kinetic rate constants 
are computed for a fixed set of parameters regarding the stochastic model 
(reacting volumes $V_{\rm c}$ and $V_{\rm s}$, and attempt frequency $\nu$). 
We have pointed out that the order of magnitude of the rates can be estimated 
from the ideal solution behavior (Eqs. (\ref{eq:konkoffest})).

There is a large amount of literature on simulation methods/studies of flexible 
or semiflexible equilibrium polymers but most of them resort to various purely 
MC schemes which sample a grand-canonical ensemble (with addition/removal of 
individual molecules homogeneously in the simulation box) subject to 
additional constraints on chemical potentials in order to impose chemical 
equilibrium,~\cite{frenkel-smit,lu-04}
or use sometimes artificial MD schemes to compute only the equilibrium static 
quantities.~\cite{lisal-06} 
Our exploration of the reactive semi-grand canonical ensemble by MD gives 
similar information on thermodynamic and structural quantities (such as 
distributions of filament length populations, various monomer/filament pair 
correlation functions etc), but in addition yields dynamic information on 
diffusive and reactive processes both at equilibrium, as illustrated in the 
present work, and potentially outside equilibrium.

Reaction-diffusion aspects of biofilaments (actin) in presence of explicit 
reactive free monomers have been studied recently in inspiring BD 
studies.~\cite{lee-08,lee-09,guo-09,guo-10}
The depolymerization steps are modeled as explicit end-monomer detachments 
directly triggered by an independent stochastic process with fixed 
depolymerizing rate. 
The polymerization step relies on the detection of free monomers entering, 
by diffusion, into a capture zone volume defined at the reactive end. 
Subsequent to its entry, the monomer is automatically treated as reactive and 
subsequently attached to the pre-existing flexible filament. 
In all these studies, the opposite polymerization and depolymerization steps do 
not satisfy micro-reversibility.
While Lee and Liu~\cite{lee-08,lee-09} studies are very specific and concerned 
with far from equilibrium network developments close to a moving disk 
(including branching and capping phenomena) to investigate the origin of disk 
motility, Guo et al.,~\cite{guo-09,guo-10} 
are concerned with more basic phenomena on single F-actin filaments regarding 
size and composition fluctuations. 
They treat a single unbounded filament surrounded by free Brownian monomers at 
a strictly constant concentration, mimicking an homogeneous stationary 
non-equilibrium situation. 
Purely ADP-actin complexes have been considered in the first study concerned 
with a stationary growth or shrinkage of the filament. 
In the next work, treadmilling has been produced by a similar BD 
study in presence of three types of actin complexes namely, 
ADP-, ADP+P$_{\rm i}$- and ATP-actin.
(Here P$_{\rm i}$ represents inorganic phosphate group). 
In these BD studies, for pragmatic reasons, chemical reactions are artificially 
accelerated with respect to a free monomer diffusive time scale of reference, 
by increasing both kinetic rates and free monomer concentration in order to 
follow diffusion and chemical events within a unique simulation time window 
allowing significant statistics on the slowest dynamical events.
Our methodology is different but works at the same mesoscopic level and 
therefore, if applied to the actin case, it would face the same wide spectrum 
of time scales problem discussed above. 
Progress on further time-space rescaling needs to be envisaged such as 
considering the elementary monomeric unit treated in the model to represent 
a collection of a few ATP-protein and/or ADP-protein complexes, with an 
effective slower diffusion constant and adapted chemical steps.

Coming back to the large classes of equilibrium and nonequilibrium 
self-assembled polymers or filaments considered in the introduction, the 
kinetic processes which are involved in their self-assembly and disassembly are 
known to be of very different nature. 
Cylindrical micelles can break anywhere and reform by end-recombinations and 
this mechanism is potentially plausible to some extent in 
biofilaments also.~\cite{bausch-11}
Biofilaments are usually subject to ATP-hydrolysis while the building 
blocks of the filaments are ATP-protein complexes which can thus be transformed 
in other complexes when integrated into filaments. 
In our method, we only restrict the chemical steps to be single unit end 
polymerization or depolymerization and that too to single monomer species. 
Our approach can however be easily extended to cover all these cases as 
illustrated in the next section where we discuss some aspects of actin modeling.

\subsection{Parametrization of our generic biofilament model to F-actin}

For illustrative purposes, we discuss below the application of our method to 
actin at the level of one free monomer particle being one single protein 
complex with ADP or ATP. 
Referring to the study of actin by Guo et al.,~\cite{guo-09,guo-10} 
we discuss in detail the parametrization needed in the equilibrium polymer case 
where any monomer, either free or integrated in a F-actin filament, remains 
permanently in the actin-ADP complex state. 
We then briefly discuss extension towards non-equilibrium actin filaments 
subject to irreversible ATP hydrolysis.

To begin with, we remark that the diffusion time of free monomers in our 
study, depending on the mass of the free monomer and various parameters of the 
MPCD solvent, turns out to be $5 \;\tau_0$. 
This can be used to set the basic time unit as $\tau_0 = 0.5 \times 10^{-6}$ s 
when the experimental Globular-actin diffusion coefficient of 
$10^{-11}$ m$^2$/s is taken into account.~\cite{mcgrath-98}
The size of the monomer is $0.5\: a_0$, fixing our length unit to $a_0=10$ nm. 
As actin has a double-stranded helix structure,~\cite{alberts} it is usually 
in simplified models as a single strand such that each monomer addition 
increases the effective contour length of the filament by half the size of a 
unit.~\cite{ranjith-09} 
This implies we should take $l_0=0.25 \;a_0$. 
The force constant $k$ is not a sensible physical parameter but it should be 
fixed to a sufficiently large value to avoid unphysical fluctuations of the 
contour length. 
The bond fluctuations are small with respect to $l_0$ if the quantity $x_0$ 
defined in Eq.~(\ref{eq:eqc}) satisfies $x_0\gg 1$. 
In the present discussion, we can keep the value $k=1600$ in 
simulation units, as this value leads to $x_0=10$. 
Next, we must fix $k'$ such that we reproduce the experimental value of the 
persistence length of the pure ADP-F-actin at $T=293$ K which is 
$l_{\rm P} = 9$ $\mu$m.~\cite{isambert-95}
This gives a value of $k'= 3.6 \times 10^{3}$ $k_{\rm B}T$.
The value of the equilibrium constant $K^0$ will next be imposed in the model by
using Eq.~(\ref{eq:eqcs}) valid as long as $x_0\gg 1$, and by exploiting 
the freedom in choosing the bonding energy parameter $\epsilon_0$. 
We thus find $\epsilon_0=18.7$ $k_{\rm B}T$ in order to get the experimental 
critical concentration of 
$\rho_{\rm c}=(K^0)^{-1}=1.8$ $\mu$M.~\cite{fujiwara-07}
These set of parameters still require the rate $\nu$ to be adjusted to 
match the experimental data of the off-rates (the on-rates are automatically 
satisfied if $K^0$ is correctly fixed). 
As one has different values at both ends, namely 
$k_{\rm off}^{\rm ba}=5.4$ s$^{-1}$ at the barbed end and 
$k_{\rm off}^{\rm po}=0.25$ s$^{-1}$ at the pointed end,~\cite{fujiwara-07} 
one gets by applying Eq.~(\ref{eqn:konkoffest-a}), 
$\nu^{\rm ba}
\approx 7\times 10^{8} $ s$^{-1}$ and 
$\nu^{\rm po}
\approx 3 \times 10^7 $ s$^{-1}$.

Moving now to F-actin filaments consisting of three different forms of 
complexes, namely ATP-actin, ADP+P$_{\rm i}$-actin and ADP-actin, requires 
associating a flag to each monomer in the model. 
Any active end of the filament (``barbed end'' or ``pointed end'') may differ 
in monomer composition (hydrolysis status) and hence specific rate constants 
for each case must be adjusted in the model parameters.  
The kinetic model for the associated irreversible hydrolysis of the ATP form 
needs also to be specified, as well as the associated kinetic rates. 
To discuss the parametrization in this context, let us adopt a simplified 
treatment~\cite{ranjith-09} where the intermediate form of ADP+P$_{\rm i}$-actin
is merged with the ADP-actin form so that only two distinct forms are 
retained and where the hydrolysis step, which irreversibly 
changes the ATP-complex in the filament into an ADP-complex, takes place only 
at the boundary of an ATP cap and pure ADP section (vectorial process). 
In such a block-copolymer, the different parameters 
$k'^{\rm ADP}$,$\epsilon_0^{\rm ADP}$ and $\nu^{\rm ADP}$ for the 
ADP block at the pointed end (and at the barbed end in case the ATP cap has 
vanished) can be adjusted on those of the pure ADP case discussed previously. 
One can proceed in the same way for the ATP block adjusting the parameter 
$k'^{\rm ATP}$ to match its stiffer character (persistence length of 
15-17 $\mu$m~\cite{gittes-93,ott-93}) and further impose values of 
$\epsilon_0^{\rm ATP}, \nu^{\rm ATP}$ for the barbed end (and the pointed end in 
case the ADP block would have vanished) in order to match the corresponding 
experimental values of kinetic rates and hence of the equilibrium constants. 
Finally, an independent stochastic process with rate fixed to its experimental 
value would convert the first ATP-complex of the cap into the ADP form.
Actin simulations would be run as explained in Section~\ref{sec:gen-model} 
but with a much large variety of reactive monomers including ATP (or ADP) 
barbed end (or pointed end) monomers, ATP monomers at the diblock boundary 
and ATP (or ADP) free monomers in the reactive volume centered on an active 
end (barbed or pointed). 
For the hydrolysis reaction, if sampled during a MD step, the reaction is 
automatically performed as an irreversible step. 
In all other cases where a monomer is selected to perform a specific reaction, 
an attempted MC move is performed which is then subject to the appropriate 
acceptance criterion.

\section{Conclusion \label{sec:conclusion}}

To conclude, we have presented a hybrid MC-MD model to simulate living 
biofilaments in the framework of the widely used wormlike chain model, with 
reactive dynamics at the ends giving rise to contour length variations. 
We believe that our mesoscopic approach, being formulated within a  
statistical mechanics framework, could be the starting point for numerous 
applications and extensions. 
The study of the growth of filament bundles in free or in confined space are in 
progress. 
More generally, in the multi-scale problem of living biofilaments dynamics, 
our methodology offers a well defined model at an intermediate level which 
should open natural bridges to more atomistic models on one side and on the 
other side, to more coarse-grained models to be based on further systematic 
and controlled rescaling procedures.
On a side note, our simulation approach could also be easily adapted 
to study polymer synthesis in good solvent by living polymerization, 
in particular, to follow the time evolution of the size 
distribution of the polymers.~\cite{das-99}

\begin{acknowledgments}
The authors are grateful to Marc Baus, Ray Kapral, P. B. Sunil Kumar and 
Jean-Louis Martiel for illuminating discussions on various aspects of the work. 
We thank Georges Destree at the ULB/VUB Centre de Calcul for efficient 
programming support. 
SR wishes to acknowledge the BRIC (Bureau des Relations Internationales et de 
la Coop\'{e}ration) of the ULB, for financial support.
\end{acknowledgments}

\appendix

\section{Statistical mechanics treatment
\label{app:stat}
}

The partition function of our semi-grand canonical reactive ensemble treats 
independent solute entities divided into a set of $N_f$ living filaments
with polymer size fluctuating through single monomer (de)polymerization between 
$i=3$ and $i=z$ and a set of $N_1$ free monomers such that the total number 
$N_t$ of monomers is fixed. It reads
\begin{equation}
Q(N_t,N_f,V,T)=
\sum_{N_1,(N_i)_{i=3,z};\{\mathfrak{C}\}}
\frac{(q_1)^{N_1}}{N_1!}
\frac{(q_3)^{N_3}}{N_3!}...
\frac{(q_z)^{N_z}}{N_z!}
\label{eq:re}
\end{equation}
where $q_i$ is the canonical partition function of a single filament of size 
$i$ or of a free monomer for $i=1$.~\cite{hill-thermody,gubbins-94} 
Owing to the fluctuating size of the filaments between their lower and upper 
limits, the sum in the partition function involves all possible arrangements 
of $N_t$ monomers which satisfy the explicit constraints, jointly denoted 
formally as $\{\mathfrak{C}\}$ in Eq.~(\ref{eq:re}),
\begin{equation}
N_1+ 3 N_3+ 4 N_4+...+z N_z-N_t=0,
\label{eq:ntot}
\end{equation}
and 
\begin{equation}
N_3+ N_4 +...+N_z-N_f=0.
\label{eq:constfil}
\end{equation}
In addition, we suppose that the state point equilibrium constant $K$ associated
to monomer attachment to/detachment from a filament is independent of the 
filament size. 

This theoretical treatment can be justified within two approaches which are 
both relevant to the interpretation of our simulation experiments involving 
interacting and reactive  filaments: 
\begin{itemize}

\item It strictly applies to the ideal solution case (subjected to constraints 
given by Eqs.~(\ref{eq:ntot}) and~(\ref{eq:constfil})) as the equilibrium constant $K^0(T)$, independent of the filament size for our model,
depends only on the temperature (see Eq.~(\ref{eq:eqcs}))). 

\item It provides an approximate treatment based on a mean-field approach 
whereby the partition function is written in terms of independent effective 
single filament partition functions $q_i$. 
The $q_i$ and hence the related equilibrium constant $K$ supposed to be 
$i$ independent, are state point dependent. 
The state point dependence of $K$ needs to be approximated to close the 
set of density equations (see Eq.~(\ref{eq:Kth})).

\end{itemize}
The recursive application of Eq.~(\ref{eq:Ki}) with $K_i=K$ gives 
$q_i=q_1^{i-3} q_3 \left(K/V\right)^{i-3}$ so that the partition 
function can be rewritten as
\begin{align}
Q(N_t,N_f,V,T)&=
q_1^{(N_t-3 N_f)}q_3^{N_f} 
\left(\frac{K}{V}\right)^{(N_t-3 N_f)} \nonumber\\
&\quad \times \sum_{N_1,(N_i)_{i=3,z};\{\mathfrak{C}\}}
\frac{1}{N_1! N_3!   ...N_z!}
\left(\frac{K}{V}\right)^{-N_1}.
\end{align}
The average number densities of monomers and filaments in the thermodynamic 
limit can be estimated~\cite{hill-thermody} by searching for the largest term 
of the partition function $Q$.  
This requires solving the constrained global minimum
\begin{subequations}
\label{eq:deriv}
\begin{align}
\frac{\partial }{\partial N_1} \ln 
\left[\frac{1}{N_1!}\left(\frac{K}{V}\right)^{-N_1}\right]-\lambda &=0, \\
\frac{\partial }{\partial N_i} 
\ln \left[\frac{1}{N_i!}\right]-i \lambda - \mu&=0\ \ (i=3,z),
\end{align}
\end{subequations}
where $\lambda$ and $\mu$ are Lagrange multipliers related respectively to 
the constraints Eqs.~(\ref{eq:ntot}) and~(\ref{eq:constfil}). 
Use of the Stirling's approximation leads to the set of equations
\begin{align}
\ln N_1 + \ln (\frac{K}{V})+ \lambda &=0, \\
\ln N_i + i \lambda - \mu&=0\ \ (i=3,z).
\end{align}
In terms of reduced number densities $\hat{\rho}=\rho K$, one gets
\begin{align}
\hat{\rho}_1 &= \exp{(-\lambda)} \\
N_i = \exp{(- [i \lambda + \mu])}&= 
\hat{\rho}_1^{i} \exp{(- \mu)}\ \ (i=3,z).
\label{eq:Ni}
\end{align}
The combination of Eqs.~(\ref{eq:constfil}) and~(\ref{eq:Ni}) gives 
$\exp{(-\mu)}=N_f \left(\sum_{i=3}^z \hat{\rho}_1^{i}\right)^{-1}$, 
hence leading to the 
filament densities Eq.~(\ref{eq:Ni1}), while the implicit equation in free 
monomer density Eq.~(\ref{eq:b1}) follows by direct substitution of filament 
densities in the constraint Eq.~(\ref{eq:ntot}).

\section{First-order estimate of equilibrium constant in the presence 
of excluded volume interactions 
\label{app:kik0}
}

The pressure expansion of a mixture of monomers and filaments in terms of 
activities reads~\cite{hill-thermody}
\begin{align}
\frac{p}{k_{\rm B}T} &=
z_1 
+ \sum_{i=3}^z z_i + b_{11}z_1^2
+ \sum_{i=3}^z b_{ii}z_i^2
+ \sum_{i=3}^z b_{1i}z_1z_i \nonumber\\
&\quad + \sum_{i=3}^{z-1} \sum_{j=i+1}^z b_{ij}z_iz_j + O(z^3),
\end{align}
where the activities are defined by
\begin{subequations}
\begin{align}
z_1&\equiv f_1 \rho_1= q_1^0 \exp(\beta \mu_1)/V,\\
z_i&\equiv f_i \rho_i= q_i^0 \exp(\beta \mu_i)/V.
\label{eq:f}
\end{align}
\end{subequations}
Here the coefficients $b_{ij}$ are related to two-body integrals by 
Eqs.~(\ref{eq:bz}) and Eqs.~(\ref{eq:bzint}).
The number densities can then be expressed as
\begin{align}
\rho_1&=z_1 \left( \frac{\partial \beta p}{\partial z_1} \right)
= z_1 ( 1 + 2 b_{11}z_1 + \sum_{i=3}^z b_{1i}z_i),\\
\rho_k&=z_k \left( \frac{\partial \beta p}{\partial z_k} \right)
= z_k ( 1 + 2 b_{kk}z_k + b_{1k}z_k
+\sum_{j=3, j \neq k}^z b_{kj}z_j).
\label{eq:density}
\end{align}
In the above equations, the series are limited to quadratic terms in $z_i$.
These relations can be inverted by writing
\begin{align}
z_1&= \rho_1 + a_{11}\rho_1^2 + \sum_{i=3}^z a_{1i}\rho_1 \rho_i\\
z_i&= \rho_i + a_{ii}\rho_i^2 + a_{1i}\rho_1\rho_i+
\sum_{k=3,k\neq i}^z a_{ki}\rho_i \rho_k
\end{align}
where the coefficients $a_{ij}$ can be obtained by substitution of the 
latter expansions in the density expansions (Eq.~(\ref{eq:density})), 
followed by term by term identification, giving
\begin{align}
a_{11} & = - 2 b_{11},\\
a_{1i} & = a_{i1} = -b_{1i},\\
a_{ii} & =-2b_{ii},\\
a_{ik} & =-b_{ki}.
\end{align}
Using Eq.~(\ref{eq:f}), one gets
\begin{align}
f_1 &= 1 - 2 b_{11}\rho_1 - \sum_{i=3}^z b_{1i}\rho_i ,\\
f_i &= 1 - 2 b_{ii}\rho_i - b_{1i}\rho_1 - \sum_{k=3, k\neq i}^z b_{ki}\rho_k,\\
f_{i-1} &= 1 - 2 b_{i-1,i-1}\rho_{i-1} - b_{1,i-1}\rho_1 
- \sum_{k=3, k\neq i}^z b_{k,i-1}\rho_k.
\end{align}
The substitution of these density expansions of the activity coefficients in the 
equilibrium constant expression (Eq.~(\ref{eqn:K})) leads to the requested 
expression (Eq.~(\ref{eq:expa})).


\providecommand{\noopsort}[1]{}\providecommand{\singleletter}[1]{#1}%
%


\end{document}